\documentclass[12pt,preprint]{aastex}

\newcommand{\lsun}{\mbox{L$_{\odot}$}}

\newcommand{\lir}{\mbox{L$_{\rm IR}$}}

\newcommand{\egant}{\mbox{EVP01}}

\newcommand{\buc}{\mbox{Paper~I}}
\newcommand{\spitzer}{\emph{Spitzer}}

\newcommand{\tmass}{\emph{2MASS}}
\newcommand{\msx}{\emph{MSX}}

\newcommand{\irs}{\emph{IRS}}
\newcommand{\HII}{H~{\sc ii}}

\shorttitle{Classification of luminous mid-IR sources in the LMC}
\shortauthors{Kastner et al.}

\begin{document}

\title{The LMC's Top 250: Classification of the Most Luminous
  Compact 8~\micron\ Sources in the Large Magellanic Cloud} 

\author{Joel H.\ Kastner\altaffilmark{1,2}, Stephen L.\
  Thorndike\altaffilmark{3}, Paul A. Romanczyk\altaffilmark{1}, Catherine\ 
  Buchanan\altaffilmark{1,4}, Bruce J.\
  Hrivnak\altaffilmark{5}, Raghvendra\
  Sahai\altaffilmark{6}, \& Michael\ Egan\altaffilmark{7}
}

\altaffiltext{1}{Center for Imaging Science, Rochester Institute of
Technology, 54 Lomb Memorial Drive, Rochester NY 14623
(jhk@cis.rit.edu)}
\altaffiltext{2}{Present affiliation: Visiting Astronomer, Laboratoire
  d'Astrophysique de Grenoble, Universit\'e Joseph Fourier --- CNRS,
  BP 53, 38041 Grenoble Cedex, France (joel.kastner@obs.ujf-grenoble.fr)}
\altaffiltext{3}{Department of Physics \& Astronomy, University of 
Rochester, Bausch \& Lomb Hall, P.O. Box 270171, Rochester, NY
14627-0171}
\altaffiltext{4}{Present affiliation: The School of Physics, The
  University of Melbourne, Victoria, 3010, Australia (clb@unimelb.edu.au)}
\altaffiltext{5}{Dept. of Physics and Astronomy, Valparaiso
University, Valparaiso, IN 46383}
\altaffiltext{6}{NASA/JPL, 4800 Oak Grove Drive, Pasadena, CA 91109}
\altaffiltext{7}{National Geospatial-Intelligence Agency,
  MS P-126, 12310 Sunrise Valley Dr., Reston, VA 20191-3449}


\begin{abstract}\ To ascertain the nature of the brightest compact
  mid-infrared sources in the Large Magellanic Cloud (LMC), we have
  applied an updated version of the Buchanan et al. (2006) 2MASS-MSX
  color classification system, which is based on the results of {\it
    Spitzer} Space Telescope spectroscopy, to a mid-infrared (8
  $\mu$m) flux-limited sample of 250 LMC objects for which 2MASS and
  MSX photometry is available. The resulting 2MASS-MSX (``JHK8'')
  color-based classifications of these sources, which constitute the
  most mid-IR-luminous objects in the LMC, were augmented,
  cross-checked, and corrected where necessary via a variety of
  independent means, such that only 47 sources retain tentative
  classifications and only 10 sources cannot be classified at all. The
  sample is found to consist primarily of carbon-rich asymptotic giant
  branch (AGB) stars ($\sim35$\%), red supergiants ($\sim$18\%), and
  compact H {\sc ii} regions ($\sim30$\%), with additional, small
  populations of oxygen-rich AGB stars ($\sim4$\%), dusty, early-type
  emission-line stars ($\sim3$\%), and foreground, O-rich AGB stars in
  the Milky Way ($\sim3$\%). The very large ratio of C-rich to O-rich
  objects among the luminous and heavily dust-enshrouded AGB stars in
  our LMC IR source sample is consistent with the hypothesis that
  carbon stars form easily in lower metallicity environments. We
  demonstrate that very luminous C-rich and O-rich AGB stars and red
  supergiants, identified here primarily on the basis of their JHK8
  colors, also appear as distinct clusters in {\it Spitzer} IRAC/MIPS
  color-color diagrams. Thus, in principle, the IRS-based IR
  photometric classification techniques applied here to the LMC can be
  applied to any external galaxy whose most luminous IR point sources
  are detectable and resolvable by 2MASS and Spitzer.
\end{abstract}


\section{INTRODUCTION} \label{sec:intro}

The very late evolutionary stages of stars of initial mass
$\stackrel{>}{\sim}1$ $M_\odot$ --- and both the very early {\it and}
very late evolutionary stages of high-mass ($\stackrel{>}{\sim}10$
$M_\odot$) stars --- are characterized by high luminosities and
obscuration by thick, dusty circumstellar envelopes that absorb
photospheric emission and re-radiate this emission strongly in the
mid- to far-infrared. Such objects should dominate the mid-infrared
point-source populations of nearby, external galaxies. Indeed, in the
era of the {\it Spitzer Space Telescope}, individual rapidly
mass-losing evolved stars and massive young stellar objects have now
been detected, {\it en masse}, throughout the Local Group (e.g.,
Blum et al.\ 2006; Jackson et al.\ 2006, 2007; Cannon et al.\ 2006).

Because of its relative proximity and its large stellar populations
located at essentially uniform distance, the Large Magellanic Cloud
(LMC) is a particularly fruitful subject for studies intended to
characterize these short-lived, IR-luminous stellar populations. {\it IRAS}
and {\it ISO} mid-IR surveys of the LMC (e.g. Loup et al.\ 1997; van Loon et
al.\ 1999a; Trams et al.\ 1999) demonstrate that the most highly
evolved (dustiest) intermediate-mass (asymptotic giant branch; AGB)
and high-mass (red supergiant; RSG) stars are plentiful, perhaps
dominant, among the luminous IR point-source membership of the
LMC. Because such rapidly mass-losing AGB and RSG stars dominate the
rate of return of nuclear-processed material to the interstellar
medium (ISM), these objects play especially important roles in the
chemical evolution of galaxies. Such stars also represent key tests of
stellar evolution theory. Even before the advent of major mid-IR space
missions, the LMC's AGB star populations served as litmus tests for
models of, e.g., shell burning and carbon star production (Iben \&
Renzini 1983).


With the publication of the initial results of the \spitzer\ (Werner
et al.\ 2004) InfraRed Array Camera (IRAC) and Multiband Imaging
Photometer for Spitzer (MIPS) imaging survey of the LMC (``Survey of
the Agents of a Galaxy's Evolution'' [SAGE]; Meixner et al.\ 2006),
broad-band infrared photometry at wavelengths of 3.6, 4.5, 5.8, 8.0
and 24 $\mu$m is now available for over $30,000$ mass-losing evolved
stars in the LMC (Blum et al.\ 2006). For most of these objects, the
SAGE IRAC/MIPS photometry can be used to distinguish between, e.g.,
RSGs and AGB stars, and even between oxygen-rich and carbon-rich AGB
stars. However, the most dust-obscured and, hence most luminous
infrared sources in the LMC cannot be readily classified --- e.g., as
C-rich vs.\ O-rich AGB stars, or even as young planetary nebulae vs.\
compact \HII\ regions associated wth massive pre-main sequence stars
--- solely on the basis of SAGE colors (Blum et al.\ 2006).

For these, the most mid-IR luminous LMC sources, two predecessor
infrared surveys --- the Two Micron All-sky Survey (\tmass) and the
LMC survey conducted by the Midcourse Space Experiment (\msx) ---
previously have provided photometry in the wavelength range
1.2--8.3~\micron. Egan et al.\ (2001; hereafter \egant) compared the A
(8.3~\micron) band magnitudes obtained from the \msx\ survey of the
LMC with J (1.25~\micron), H (1.65~\micron), and K (2.17~\micron)
magnitudes obtained from \tmass. They identified 11 categories of
stellar populations among the resulting sample of 1664 objects.
\egant\ cross-checked their classifications of objects believed to
reside in the LMC using spectral type data obtained from the SIMBAD
database; however, very few of the most IR-luminous (therefore most
highly optically obscured) objects had classifications listed
in SIMBAD. 

The use of IR spectra alleviates most of the ambiguity that results
from assigning stellar classes based on IR photometry alone
(e.g., \citep{gro95, van98, tra99}).  The spectroscopic
validation and refinement of infrared-color-based classification systems
necessary to interpret the large volume of photometric data
flowing out of SAGE and other \spitzer\ imaging surveys of
Local Group galaxies is made possible by the \spitzer\
Infrared Spectrograph (\irs; Houck et al.\ 2004\footnote{The IRS was a
collaborative venture between Cornell University and Ball
Aerospace Corporation funded by NASA through the Jet
Propulsion Laboratory and Ames Research Center.}). Several
{\it Spitzer} IRS surveys of the LMC have been carried out,
most of these aimed at its evolved star populations
(e.g., Buchanan et al.\ 2006; 
Zijlstra et al.\ 2006; Groenewegen et al.\ 2007; Matsuura et
al.\ 2006; Speck et al.\ 2006; Stanghellini et al.\
2007). 

Buchanan e al.\ (2006), hereafter \buc, conducted a \spitzer\
spectroscopic study of a sample of the most luminous 8~\micron\
sources in the LMC. These sources were chosen from the \egant\
\tmass-\msx\ catalog with the expectation (indeed, bias) that the
sample would be dominated by highly evolved, rapidly mass-losing
stars.  The \irs\ spectra covered the range 5 to 35~\micron, allowing
the determination of evolved star envelope chemistries through
identification of spectral signatures of C-rich and O-rich dust.  The
IR luminosities and evolutionary status of the objects were derived
from the combined \tmass, \irs\ and (where available) {\it IRAS}
spectral energy distributions.

Among the sample of $\sim$60 objects studied in Paper I, we identified
16 C-rich AGB stars, 4 O-rich AGB stars, 21 RSG stars, and 2 OH/IR
stars (one supergiant and one AGB star).  This sample also included 11
\HII\ regions --- presumably very young O or early B stars that are
deeply embedded in their nascent, dusty, molecular clouds --- as well
as 2 B supergiants with peculiar mid-IR spectra suggestive of the
presence of circumstellar disks (Kastner et al.\ 2006).  We were thus
able to establish a revised set of classifications of luminous 8
$\mu$m LMC sourses on the basis of the \irs\ spectra, IR spectral
energy distributions, and IR luminosities of the Paper I sources.  In
particular, all of the objects \egant\ classified as PNe were
reclassified as \HII\ regions; most objects classified by \egant\ as
OH/IR stars were reclassified as C-rich AGB stars; and objects
classified by \egant\ as O-rich AGB stars were reclassified as either
RSGs or (in a handful of cases) foreground, mass-losing, O-rich AGB
stars (Mira variables) in the halo of the Milky Way. These results led
to a revised \tmass-\msx\ ($J, H, K, A$ band, i.e., 1.25, 1.65, 2.2,
8.3 $\mu$m) color-color and color-magnitude classification scheme ---
hereafter referred to as the ``JHK8'' scheme --- for luminous 8 $\mu$m
sources in the LMC (\buc). The JHK8 color-color classication scheme, being
distance-independent, is in principle applicable to any studies of
Galactic or extragalactic infrared source populations that make use of
combined near-infrared and 8 $\mu$m photometry.

The \buc\ LMC \irs\ sample was selected to cover representative
subsets of the types of luminous 8~\micron\ sources in the LMC, but
not their relative numbers. Therefore that study is not proportionally
representative of the luminous mid-IR stellar populations of the
LMC. Here, we revisit the entire sample of luminous 8~\micron\ sources
in the \egant\ lists to reclassify these objects on the basis of the
\buc\ JHK8 color criteria (revised so as to account for updates to the
\tmass\ and \msx\ photometry), as well as information available in the
literature and in the {\it Spitzer} IRS archive. We thereby determine
the distribution of infrared spectral types among a complete, 8 $\mu$m
flux-limited sample of infrared sources, so as to elucidate the
natures of the most mid-IR-luminous objects in the LMC. We then examine
these results in light of SAGE point-source photometry
obtained for the most luminous mid-IR sources in the LMC.

\section{THE SAMPLE} \label{sec:obs}

We began with all 1664 objects from the sample compiled by \egant\
that were identified in both the \msx\ and \tmass\ surveys.  
In order to study the same population sampled for the \buc\
Spitzer/IRS spectral study, we apply the same 8.3~\micron\ magnitude
limit, A$ \le 6.5$ or $F_{8.3} \ge 150$~mJy (based on the \msx\
photometric calibration cited in EVP01). Imposition of this limit
eliminates $\sim$70\% of the \egant\ sample.  Objects that \egant\
classified as main sequence stars (``MS(V)'') were then discarded, on
the grounds that these objects are neither evolved nor in the LMC.
Those objects that \egant\ classified as MIII and A-K III and have
$J\stackrel{<}{\sim}7.5$ were assumed to be giants residing in the
halo of the Milky Way and were therefore also rejected.  The objects
MSX LMC 140, 421, 946, 1046, 1080, 1270, 1419, 1734, 1752, 1015, and
1049 were not designated as either M III or A-K III stars by \egant\
but are classified as ``star'' or ``M star'' in the SIMBAD database.
Given that the locations of these objects within 2MASS/MSX color-color
diagrams are indicative of photospheric rather than dust emission,
these objects are also most likely foreground, first-ascent red giants
and were thus discarded. In addition, on the basis of updated 2MASS
positions, we find that the \egant\ sources MSX LMC 470, 505, 585, and
1107 likely had spurious 2MASS associations.

The final sample considered here then consists of 250 objects (Table
\ref{tab:objtable}). As the \msx\ and \tmass\ photometric catalogs
have been revised since publication of \egant, we used IRSA's
\verb+gator+
tool\footnote{http://irsa.ipac.caltech.edu/applications/Gator/} to
compile the most recent available (2007 June) \msx\ and \tmass\ photometry for
these 250 sources. We find that the (version 6) \msx\ A band
magnitudes presently available via \verb+gator+ are typically
$\sim30-50$\% brighter than the (version 5) A band magnitudes listed
in \egant; this discrepancy does not affect the fundamental results
presented here (see below and \S 5). 
For the vast majority of
the \tmass\ data, the changes are on the order of $\sim1$\% or less,
the exceptions being the small fraction of Table \ref{tab:objtable}
objects for which the \tmass\ data available to EVP01 evidently
suffered from source confusion or data reduction problems.

Although we believe Table 1 to represent the LMC's most luminous,
compact mid-IR sources, there remains some uncertainty in the
assignment of individual objects to this sample. Given the $6''$ \msx\
A-band point spread function (PSF) there is the possibility that some
\msx\ sources --- particularly those associated with compact \HII\
regions, i.e., in clusters or regions of star formation --- have been
associated with the wrong \tmass\ source or are contaminated with flux
from nearby sources (the identification of 2MASS sources with MSX
sources is discussed in detail in EVP01). Perhaps more importantly,
because AGB stars (many of which are Mira variables) can exhibit
strong IR variability, some sources included in (or excluded from)
this study may actually exhibit time-averaged 8 $\mu$m fluxes that are
somewhat below (or above) the cutoff we have imposed. However, we do
not expect these quasi-random source inclusions (exclusions) to affect
the basic source population results described in \S 4.

Table \ref{tab:objtable} lists the following for the sample of 250
objects: \msx\ LMC object number (column 1); SIMBAD\footnote{simbad.u-strasbg.fr/sim-fid.pl} designated name
and spectral or object type (cols.\ 2, 3); \tmass\ J, H, K magnitudes
(with flux upper limits indicated by ``u'') and \msx\ A magnitudes, as
obtained from the IRSA databases via \verb+gator+ (cols.\ 4--7); IR
classes as determined from \tmass-\msx colors by \egant\ (col.\ 8); IR
classes determined from the spectroscopic study presented in \buc\
(col.\ 9); classifications we have assigned the objects in this paper,
first based solely on revised \buc\ color-color criteria, as described
in \S\S 3.1. and 3.2 (col.\ 10), and then
incorporating additional information gleaned from other sources, as
described in \S 3.2.1 (col.\ 11, with references listed in col.\ 13);
and luminosities of the C-rich AGB and RSG objects, based on
application of ``bolometric corrections'' developed in \buc\ (column
12; see \S 3.3). Table \ref{tab:summarytable} contains a summary of
the classifications of the Table \ref{tab:objtable} objects as listed
in column 11 of that Table.  Columns 10--12 of Table
\ref{tab:objtable} and the summary in Table \ref{tab:summarytable}
thus reflect the results of the classification analysis carried out
here (\S 3).


\section{SPITZER/IRS-BASED COLOR-COLOR AND
  COLOR-MAGNITUDE CLASSIFICATION OF LMC INFRARED
  SOURCES} \label{subsec:dis_cols} 

\subsection{Revisions to JHK8 Color-color Classification Criteria}

\begin{figure}
\epsscale{0.75}
\plotone{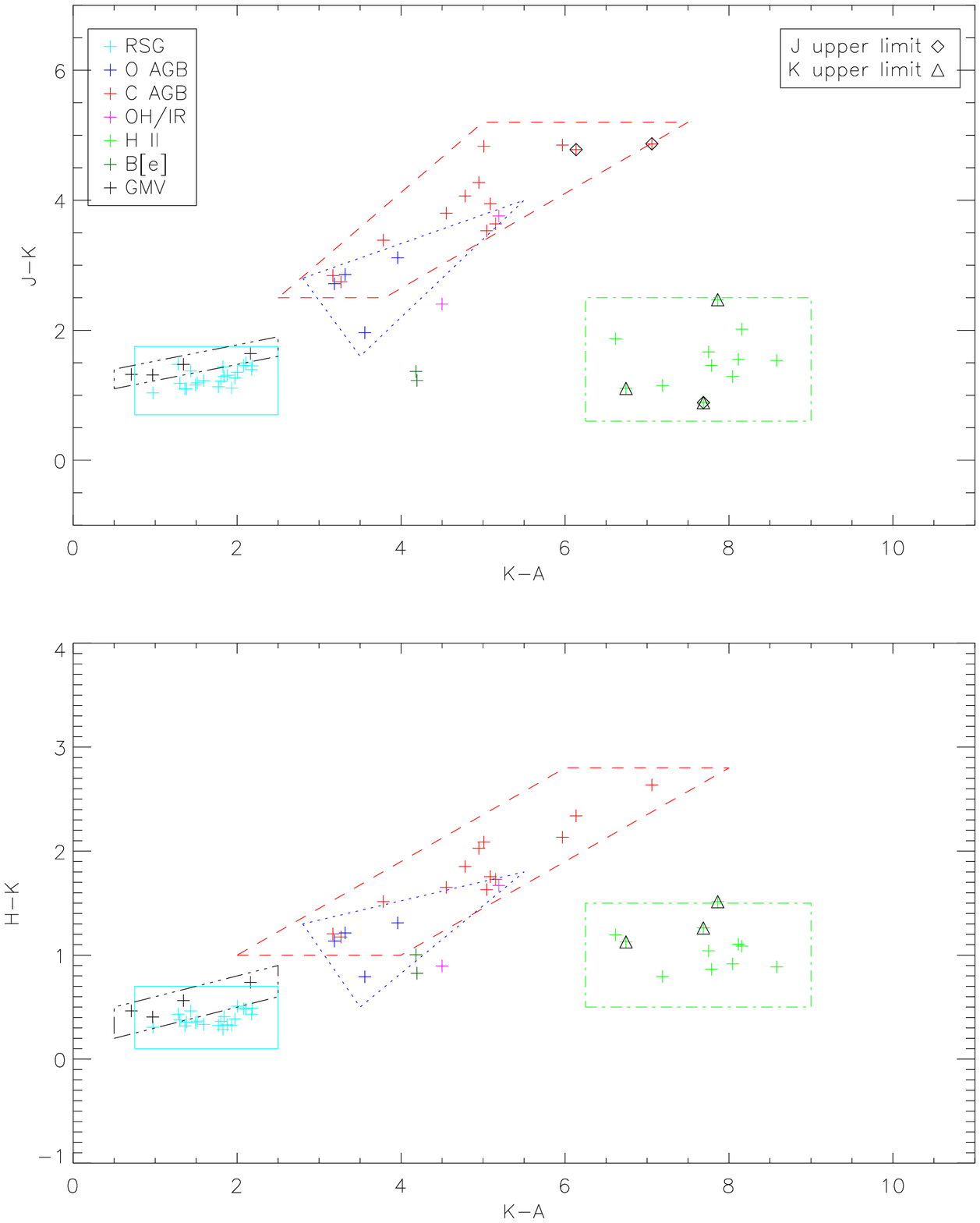}
\caption{\tmass--\msx color-color diagrams for those Table 1 objects
  studied spectroscopically with {\it Spitzer}/IRS (\buc), overlaid with revised
  object classification ``boxes'' (see text and Table 3). Objects for
  which one or more \tmass\ magnitudes are upper limits are indicated.}
\label{fig:newBucColCol}
\end{figure}

Color-color diagrams constructed
from the Table 1 \tmass-\msx\ photometry for the 57 objects studied
via IRS spectroscopy in \buc\ are displayed in
Figure~\ref{fig:newBucColCol}. As noted in \buc, these $J-K$ vs.\ $K-A$
and $H-K$ vs.\ $K-A$ diagrams, as ``calibrated'' by {\it Spitzer}/IRS,
illustrate that the RSGs, O-rich AGB stars, and C-rich AGB stars form
a sequence of increasing redness. The Galactic AGB star group
generically referred to in \buc\ as galactic Mira variables (GMVs)
occupies a subset of the RSG space and, therefore, GMVs are
indistinguishable from RSGs based solely on \tmass-\msx\ color-color
diagrams (we caution that it remains to confirm, via measurement of
their V-band amplitudes, the Mira status of some of these
``GMVs''). \HII\ regions appear as a distinct group, due to their
combination of blue $J-K$ and $H-K$ colors and very red $K-A$ colors
(this region of color-color space may also include planetary nebulae,
post-AGB stars, and/or other emission-line objects).

In light of the revisions to the \msx\ A-band photometry since
publication of \egant, and given that many compact \HII\ regions
apparently lurk among the LMC's most luminous 8 $\mu$m point sources
(\buc), we have revisited the \buc\ Spitzer/IRS-based JHK8 color-color
classification criteria.  Our revised JHK8 classification regions,
based on the positions of these source classes, are overlaid in
Figure~\ref{fig:newBucColCol} and are summarized in Table 3. The main
differences between these JHK8 color-color classification criteria and
the original \buc\ criteria are: (a) the positions of most of the
classification regions are shifted $\sim0.3$ mag redward in $K-A$
(i.e., rightward) relative to their positions in \buc, as a
consequence of the systematically larger A band fluxes of the \buc\
objects relative to those published in \egant; (b) there is
considerably more overlap between the ``O AGB'' (O-rich AGB) and ``C
AGB'' (C-rich AGB) classification regions, mainly due to our inclusion
here of the OH/IR star \msx\ LMC 807 in the ``O AGB'' group; and (c)
we expand the \HII\ region classification zone, to compensate for the
fact that the \buc\ selection criteria for such objects were overly
restrictive (\S 4.1). The O AGB region is in fact very poorly
constrained, due to the paucity of O-rich AGB stars in the \buc\
sample. Only three objects in the \buc\ study then remain outside the
regions indicated in Figure~\ref{fig:newBucColCol}. This ``outlier''
group consists of the OH/IR supergiant \msx\ LMC 1182, which is found
redward (in $K-A$) of the O-rich and C-rich AGB color-color regions,
and two B[e] supergiants with dusty circumstellar disks (\msx\ LMC 890
and 1326; Kastner et al.\ 2006) that lie between the RSG and \HII\
region boxes.

\subsection{Application of JHK8 Color-based Classification Criteria}

\begin{figure}
\epsscale{0.75}
\rotate
\plotone{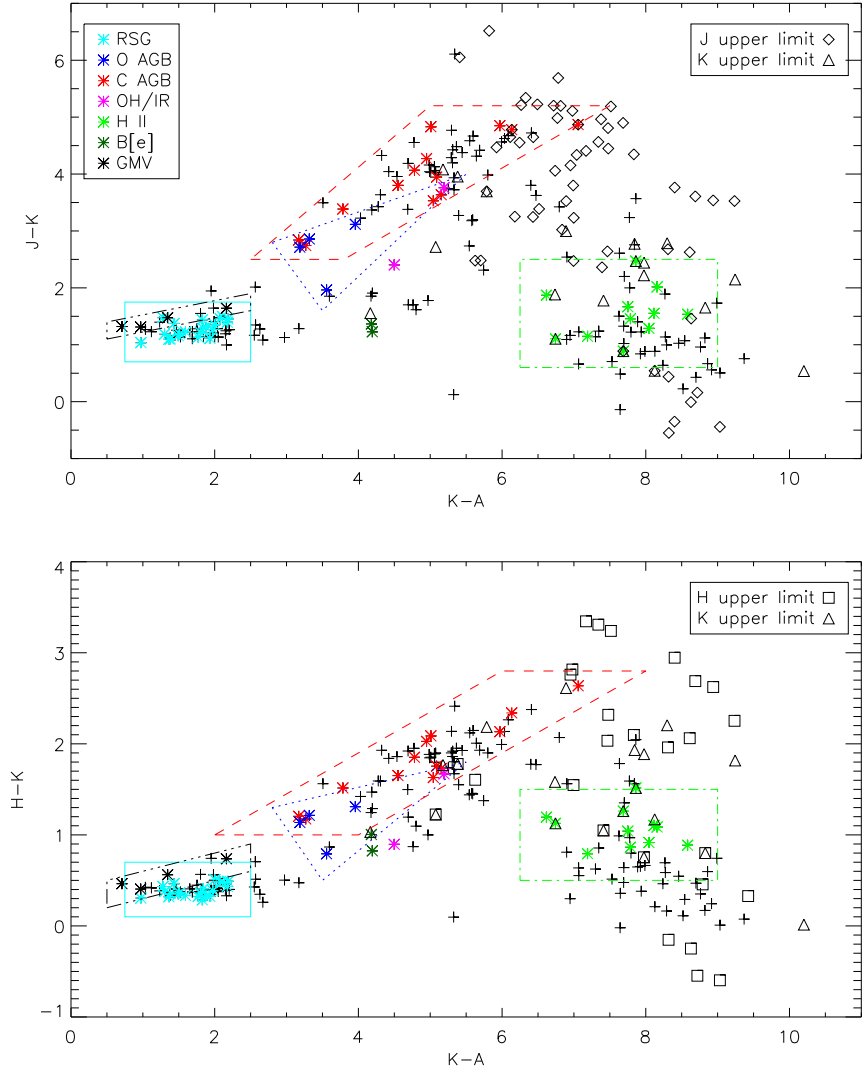}
\caption{ \protect\tmass/\protect\msx\ color-color diagrams
  displaying $J-K$ vs. $K-A$ (top) and $H-K$ vs. $K-A$ (bottom) colors
  for the Table 1 sample. The boxes indicate the revised (Table 3)
  \buc\ Spitzer IRS-based infrared color criteria for classification
  of IR-luminous LMC objects. The colored symbols
  indicate objects with IRS spectra obtained in the \buc\
  survey, with classifications as indicated in the figure
  legend. Crosses indicate objects with measured JHK magnitudes that
  were not included in the \buc\ spectroscopic study. Objects for
  which one or more \tmass\ magnitudes are upper limits are also indicated.
\label{fig:colors}} 
\end{figure}

The \tmass\ and \msx\ color-color diagrams constructed for our sample
of objects, overlaid with revised (Table 3) Spitzer/IRS-based JHK8
color classification regions, are displayed in Figure~\ref{fig:colors}.
It is apparent from Figure~\ref{fig:colors} that most objects fall
within a unique classification region. We consider these objects to be
classified with relatively high confidence. Many other objects appear
in a given classification region on only one of the color-color plots,
and we ascribe somewhat lower confidence to their classifications
(these tentative classifications are indicated by a colon in Table
\ref{tab:objtable}). Hence, the majority of Table 1 objects can be
classified (or tentatively classified) solely on the basis of their
JHK8 colors (column 10 of Table 1).

Nonetheless, because our sample is greatly enlarged over that studied
in \buc, a substantial fraction ($\sim25$\%) of Table 1 objects lack
JHK8 (color-based) classifications, even after relaxing the
classification criteria for \HII\ regions. Many others ---
particularly those with an upper limit in one or more \tmass\ bands
--- lie on or just outside of the lines defining the various
classification regions. Some objects lie within zones of confusion
between classes. For all of these sources the classifications listed
in column 10 are uncertain, ambiguous, or both. These tentative
classifications are indicated by colons. The classifications of a few
objects whose colors place them very near (within $\sim$0.2 mag of),
but outside of, the Table 3 classification regions are also flagged
with colons, while those within zones of confusion between RSGs and
GMVs or between O-rich and C-rich AGB stars are classified as
``RSG/GMV'' or ``C/O AGB'', respectively, in column 10 of Table 1.

\subsubsection{Cross-checks of JHK8 color-based classifications}

We have employed a variety of means to cross-check the JHK8
color-based classifications in column 10 of Table 1 so as to verify or
revise these classifications, as well as to determine the nature of as
many as possible of the unclassified, tentatively classified, or
ambiguously classified sources. 

1) {\it Previous classifications:} We cross-checked the JHK8
classifications against the available literature as well as the
available SIMBAD object
class and/or spectral type information (see reference list in the
notes to Table 1). Where a conflict exists between the JHK8
color-based classification and a previous 
classification, we defer to the previous (literature-based) classification.

2) {\it Color-magnitude-based classifications of red supergiants and
  Milky Way AGB stars:} Most objects in the region of overlap between
RSG and GMV classes wthin the \tmass-\msx\ color-color diagrams
(listed as ``RSG/GMV'' in Table 1) can be firmly classified on the
basis of \tmass-\msx\ color-magnitude diagrams. In
Figure~\ref{fig:magcolors}, we present such diagrams for the Table 1
sample (objects for which J and/or K fluxes are upper limits are not
displayed). As noted in \buc, LMC RSGs and galactic Mira variables are
generally readily distinguished in these diagrams, due to the relative
proximity of the latter group (\S 4.2). 
Objects for which this cross-check yielded a clear
classification of an object as ``RSG'' or ``GMV'' are indicated by
footnote ``g'' in Table 1.

3) {\it Candidate \HII\ regions included in or excluded from the SAGE
  PSC:} A check on our classification of Table 1 objects as \HII\
regions is provided by their presence --- or, indeed, lack thereof ---
in the SAGE Point Source Catalog (PSC). Since the {\it Spitzer}/IRAC
PSF is $\sim2.5''$ at 8.0 $\mu$m, many Table 1
sources (all of which appeared point-like to MSX, with its PSF FWHM of
$\sim6''$ at 8.3 $\mu$m) are resolved by IRAC and, hence, are not
included in the SAGE PSC (see \S 5). We regard all such Table 1
sources as having their \HII\ region status confirmed. On the other
hand, the minority of sources with JHK8 colors indicative of \HII\
regions that {\it do} appear in the SAGE PSC (excluding \msx\ LMC
1794, which was studied in \buc) are considered as tentative
classifications. Their inclusion in the SAGE PSC assures that these
sources are generally more compact than the \HII\ regions included in
the \buc\ spectroscopic study (most of which were well resolved by the
short-wavelength spectrometer modules of the IRS). Therefore some of
the Table 1 sources tentatively classified as \HII\ regions may in
fact be PNs, pre-PNs, or circumstellar dust disks or envelopes
associated with massive, early-type, emission-line stars (see \S\S
4.1, 4.4).

4) {\it IRAS fluxes of candidate \HII\ regions:} {\it IRAS} data
provide an additional indication of \HII\ region status, although we
regard {\it IRAS} fluxes as a rather weak constraint in this regard
given the likelihood such data may suffer from confusion in the
crowded fields typical of \HII\ regions.  We find 16 of 17 candidate
\HII\ regions in Table 1 for which {\it IRAS} data are available have
25 $\mu$m flux densities $>1.3$ Jy --- similar to or, in many cases,
larger than those of the \HII\ regions studied in Paper I.  All of
these sources display steeply rising SEDs in the {\it IRAS} data,
leading to extrapolated luminosities $\stackrel{>}{\sim}10^5$
$L_\odot$, too large for PNs. Only three of these sources (MSX LMC
502, 932, and 1186) are compact enough to be included in the SAGE PSC
(see \S 3.2.1) and hence retain tentative \HII\ region
classifications.

5) {\it Archival IRS spectra:} Last but not least, good-quality
unpublished archival {\it Spitzer} IRS spectra have become publicly
available for several Table 1 objects. Where possible, we have used
these spectra to confirm or revise source classifications (these
classifications are indicated in Table 1 by footnote {\it i}). We
defer publication and descriptions of these data
to a later paper that will be devoted to all IRS spectra of Table 1
objects obtained subsequent to the \buc\ study (Buchanan et al., in
preparation).

After applying these (five) independent cross-checks, we are able to
classify or tentatively classify 44 Table 1 objects whose classes
could not be determined on the basis of JHK8 colors (excluding objects
studied in \buc). We firmly classify another 13 objects for which the
color-based classifications are either ambiguous or uncertain. In all,
there are 13 cases in which we find that the JHK8 color-based
classifications are incorrect, and 15 in which these color-based
classifications (either tentative or secure) are supported.

\subsubsection{JHK8 Classifications: Results and ``Success Rate''}

The results of application of JHK8 color-based classifications, and
the cross-checks of these classifications, are listed in column 11 of
Table 1. Tentative classifications are indicated by ``?''. Most of
these sources either have tentative JHK8 color-based classifications
which we are unable to confirm via the cross-checks described in \S
3.2.1, or reside within the boundaries of a JHK8 classification region
for which the color-based classification is rendered questionable by
one or more cross-checks.

\begin{figure}
\includegraphics[scale=0.5,angle=90]{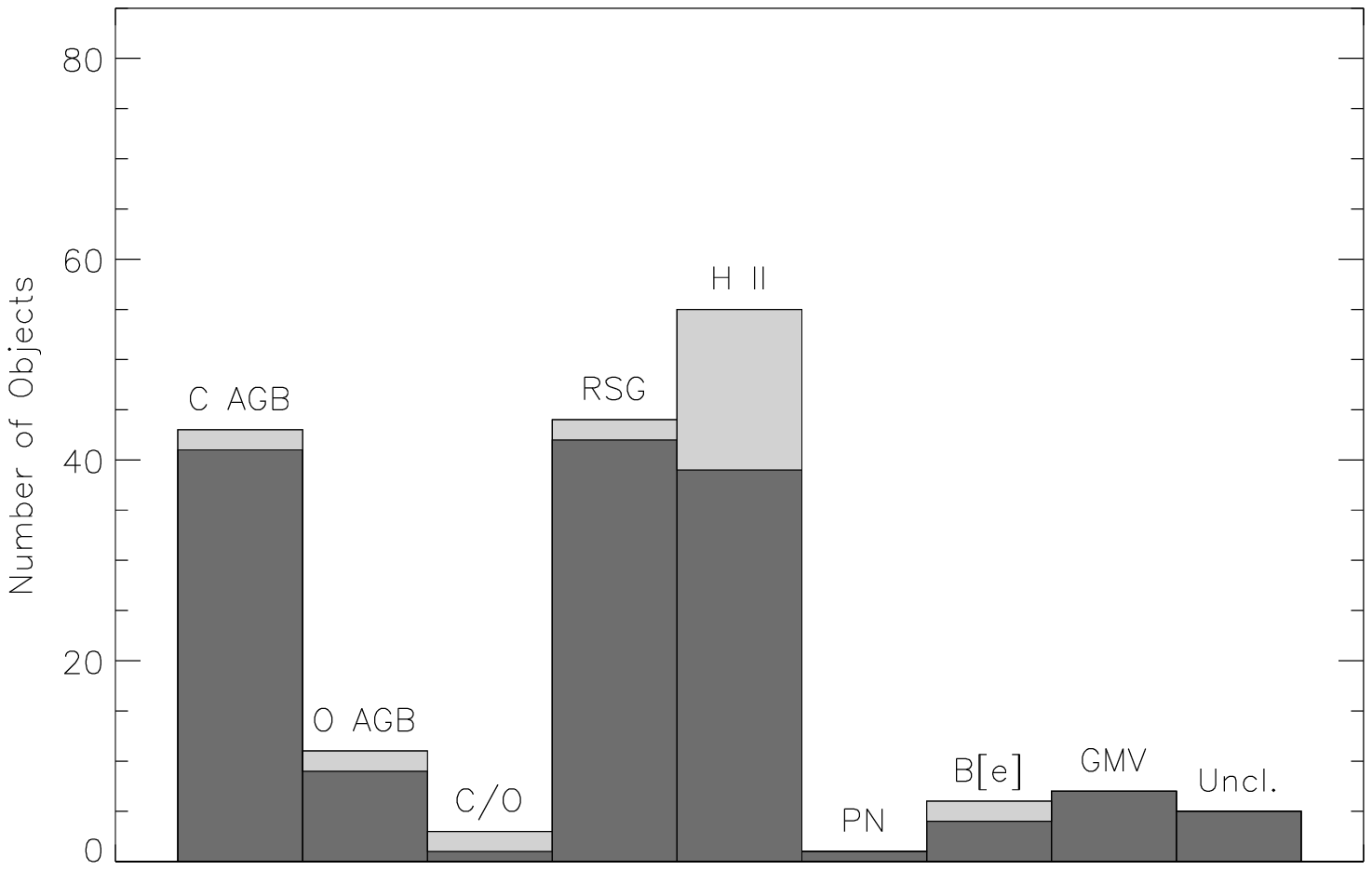}
\includegraphics[scale=0.5,angle=90]{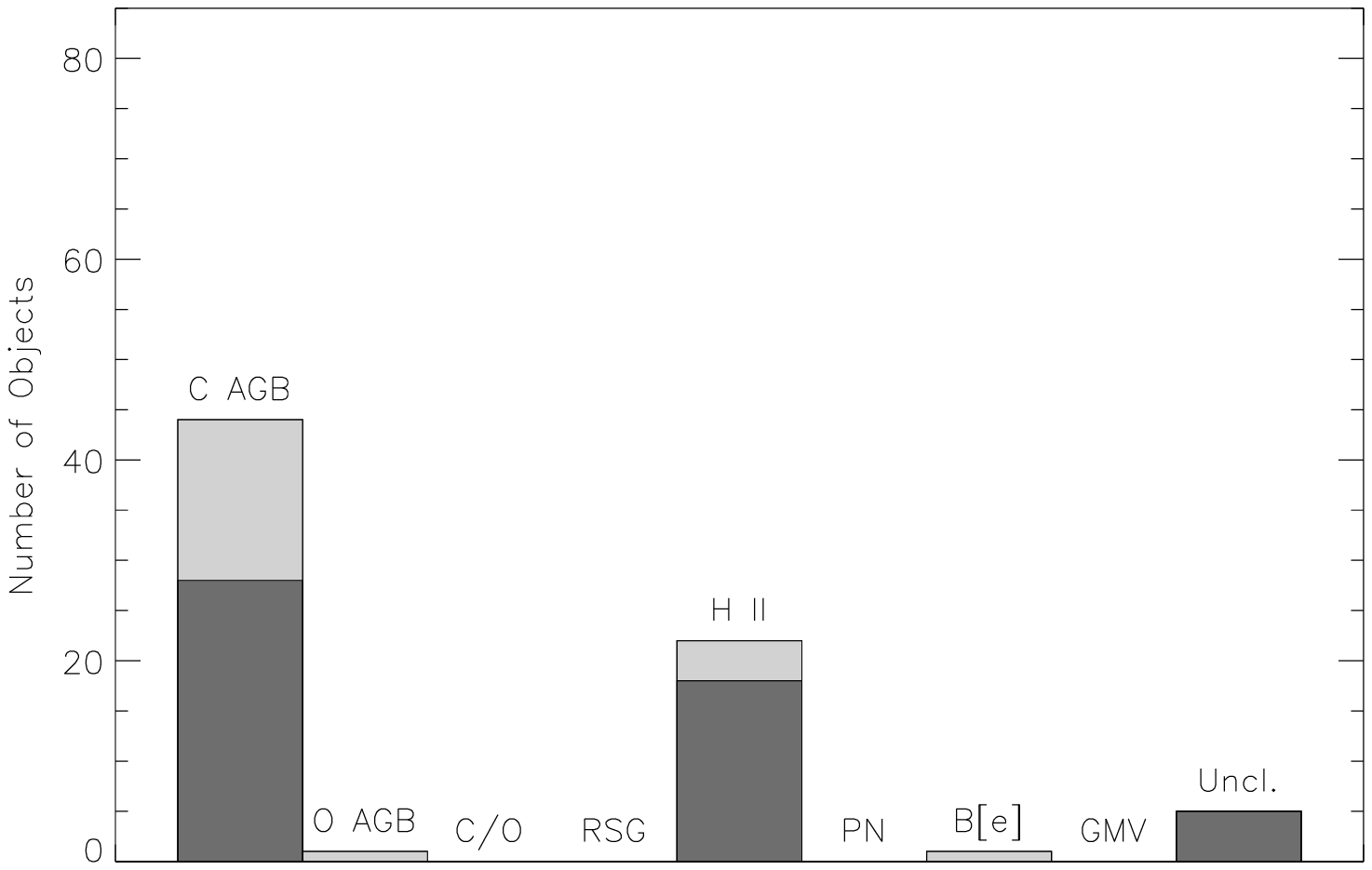}
\includegraphics[scale=0.5,angle=90]{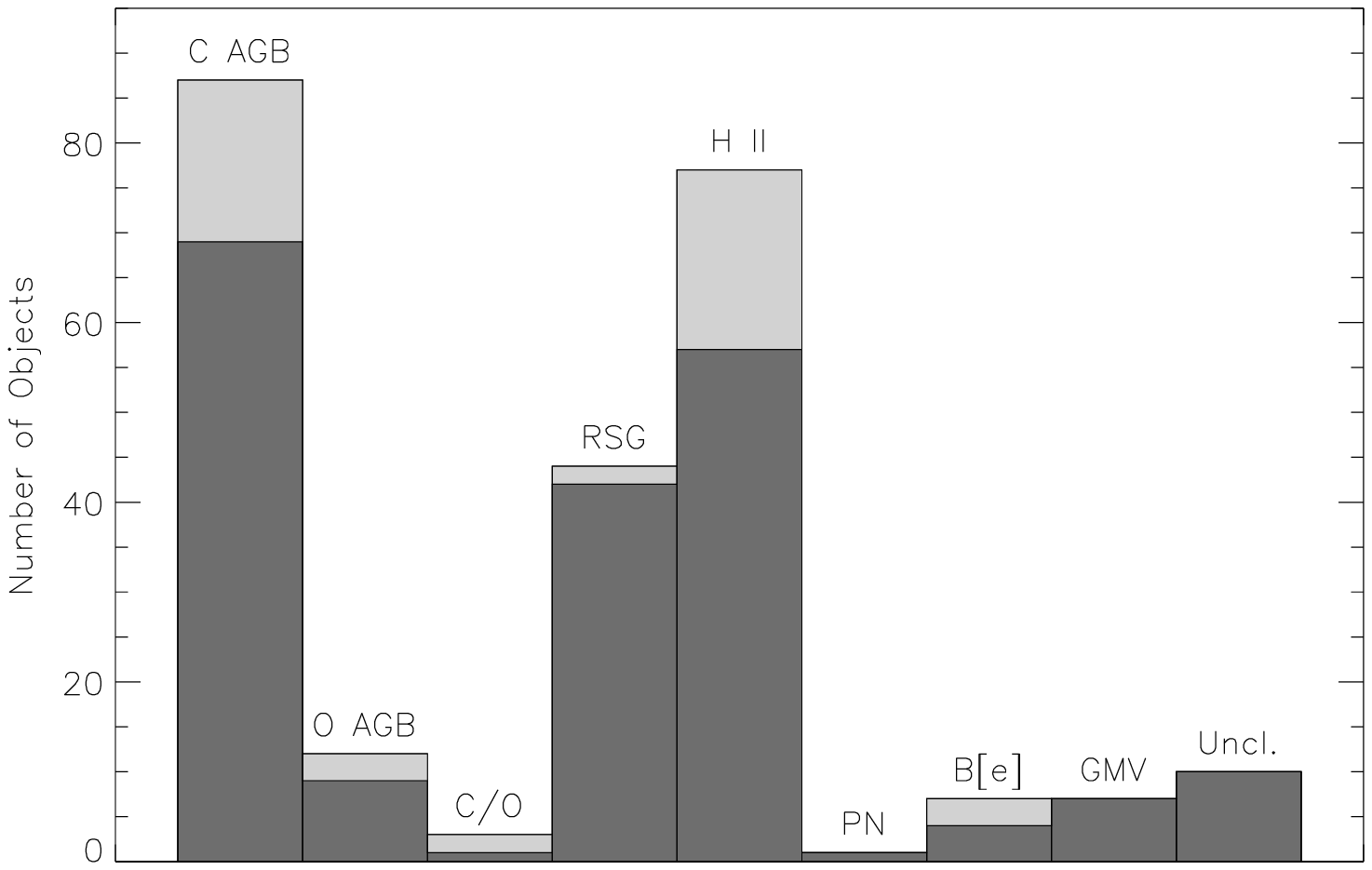}
\caption{Histograms illustrating the number of objects in each class,
  based on classifications in column 11 of Table 1, for objects with
  measured fluxes in all three \tmass\ bands (upper left), those with upper
  limits in one or more \tmass\ bands (upper right), and the entire sample
  (bottom). In each panel, the dark grey bars 
  indicate the number of objects classified with high
  confidence, while the light grey indicates the total number of
  objects in each class when tentative classifications are
  included. }
\label{fig:histogram}
\end{figure}

We display color-color diagrams
illustrating the Table 1, column 11 classification results in
Figure~\ref{fig:colorJHKA}. The final tabulations of the population
within each object class (Table 2 and Fig.\ \ref{fig:histogram}) are
based on these results. The population histograms (Fig.\
\ref{fig:histogram}) indicate that the vast majority of objects with
upper limits in one or more \tmass\ bands are either (confirmed or
candidate) C-rich AGB stars or \HII\ regions, or are unclassified. The
classification results, including individual objects of interest, are
discussed in detail in \S 4.

Comparison of columns 10 and 11 of Table 1 can be used to estimate
(albeit rather crudely) the ``success rate'' of the application of the
JHK8 color-based classification criteria in Table 3.  In
addition to the 28 cases of ``success and failure'' in applying the
Table 3 criteria to objects not included in the \buc\ sample mentioned
above, these
criteria (augmented by \tmass-\msx\ color-magnitude diagrams) also, by
definition, unambiguously and correctly classify $\sim85$\% of the 57
\buc\ sources --- the exceptions being the half-dozen or so sources
that lie in the region of overlap between the C-rich and O-rich AGB
stars, the OH/IR supergiant \msx\ LMC 1182, and the 2 \msx\ sources
associated with B[e] supergiants. Hence the JHK8 color-based
classification system recovers the correct classification for 
$\sim60$ of the $\sim90$ sources for which unambiguous spectral
classifications are available. The ``success rate'' of the Table 3
classification system thereby can be conservatively placed at
$\sim70$\%. The actual rate is likely to be higher, given that the
\buc\ sample does not proportionately represent the (relatively easily
classified) C AGB and RSG populations.

\subsection{Estimates of infrared luminosities for RSGs
and carbon stars}

For LMC RSGs and luminous carbon stars, \buc\ established empirical
relationships between integrated infrared luminosity ($L_{IR}$) and
K-band flux density. In the case of carbon stars, the relationship
depends on $K-A$ color, while for RSGs the ratio of K-band flux to
$L_{IR}$ is roughly constant over the observed (relatively small)
range of RSG $K-A$ colors. We have applied these ``bolometric
corrections'' to our sample; the results are listed in column 12 of
Table \ref{tab:objtable}. The tabulated luminosities of the C-rich AGB
and RSG stars have estimated uncertainties of $\sim30-50$\% and 15\%,
respectively, mainly due to source variability (\buc). The carbon star
luminosities also may be systematically overestimated by $\sim30$\%, as
a consequence of systematic errors in MSX A-band magnitudes (\S 5).
Because the Table 1 sample is biased toward the highest-luminosity
mid-IR sources in the LMC, the estimated values of $L_{IR}$ provide an
indication of the peak luminosities reached by LMC carbon stars and
RSGs (\S 6).


\section{DISCUSSION} \label{sec:dis}

\subsection{\HII\ Regions} \label{subsec:dis_class_pn}

We identify 77 sources ($\sim30$\% of the sample) as candidate \HII\
regions. A substantial fraction of these sources fall outside of the
original \buc\ \HII\ region classification ``boxes'' in one or both of
the \tmass-\msx\ color-color diagrams (Fig.~\ref{fig:colors}), even
after accounting for updated \tmass-\msx\ photometry, because the
\HII\ regions studied spectroscopically in \buc\ were selected so as
to have a fairly narrow range of \tmass-\msx colors resembling those
of PNe. In addition, a substantial number of the sources in this
region of JHK8 color-color space have upper limits in one or more
\tmass\ bands. Nevertheless, after applying ``expanded'' \HII\ region
classification criteria (Table 3) and cross-checking with the SAGE PSC
(\S 3.2.1), most are confirmed to be \HII\ regions. Only $\sim25$\% of
the \HII\ region classifications (19 objects) in Table 1 remain tentative.


Some of these objects may be PNs or pre-PNs. As noted in Paper I, many of
the objects classified as \HII\ regions are difficult to distinguish
from planetary nebulae purely on the basis of \tmass-\msx
colors. Indeed, of the objects that fall within the \HII\ region box
on both of the color-color diagrams, all but 2 objects (\msx\ LMC 358
and 360) were previously classified as PNe by \egant.  However, all 11
of the \egant\ PN candidates selected for study in \buc\ were revealed
to be \HII\ regions associated with young OB stars. These
reclassifications were based on {\it Spitzer} IRS spectral features,
extensive surrounding nebulosity apparent in optical/IR images, large
mid-infrared source dimensions as inferred from IRS data, and/or
exceedingly large luminosities ($\stackrel{>}{\sim}10^5$ $L_\odot$).

Furthermore, our search of the literature reveals
that the Table 1 sample includes only 1 PN, and this object, MSX LMC
561 (SMP 69, Sanduleak et al.\ 1978; see also Shaw et al. 2006), has
the JHK8 colors of a C-rich AGB star.
Hence, we are reasonably confident of the accuracy of the
classifications of the majority of the candidate (tentatively
classified) \HII\ regions in Table 1.  Medium-resolution mid-infrared
spectroscopy and close examination of archival imagery likely would
establish more firmly the natures of these objects.

\subsection{Red supergiants and oxygen-rich AGB stars}
\label{subsec:dis_class_o} 

A total of 56 objects in Table 1 ($22$\% of the sample) are identified
(51) or tentatively identified (5) as oxygen-rich, mass-losing evolved
stars.  Of these, the vast majority (44 objects; $18$\% of the sample) are
confirmed (42) or candidate (2) RSGs, most of which fall in the RSG
classification box in both color-color diagrams.  Only 12 confirmed
(9) and candidate (3) O-rich AGB stars appear to be present in our
sample. The total number of ``O AGB'' stars may be as high as 19,
however, depending on the nature of the 3 stars with ambiguous ``C/O
AGB'' classifications as well as 4 apparently ``underluminous'' RSGs
(see below).

Included in the O-rich evolved star populations are \msx\ LMC 807 and
1171 (classified ``O AGB'' and ``O AGB?'', respectively) and \msx\ LMC
1182 (``RSG''). These objects are OH/IR stars whose mid-IR silicate
features are partially in absorption (\buc; \S 3.2.1). The last object
(IRAS 04553$-$6825) is particularly well-studied, being the best known
LMC analog to the class of Galactic OH/IR supergiants typified by NML
Cyg (e.g., van Loon et al.\ 1998a). Indeed, the IRS spectrum of \msx\
LMC 1182 (\buc) bears an uncanny resemblance to the archival {\it ISO}
spectrum of NML Cyg (Kastner et al., in prep.).

Four stars here classified or tentatively classified as RSGs on the
basis of 2MASS-MSX colors, \msx\ LMC 506, 551, 591, and 813,
have inferred bolometric luminosities of $\sim3-6\times10^4$
$L_\odot$. Such luminosities are more consistent with bright AGB than
RSG status. If these objects are indeed O-rich AGB stars (as opposed
to very dusty RSGs), this would suggest that the large ratio of C-rich
to O-rich AGB stars in our sample (see \S 4.3) may be in part a
selection effect resulting from our minimum \msx\ A-band flux
criterion combined with the ``blue'' mid-IR SEDs that are characteristic of
O-rich AGB stars even at relatively high luminosity
and mass-loss rate. Note that all three stars have A-band
magnitudes not far from our cutoff of A$\sim6.5$. Hence, infrared
spectroscopic observations of fainter O-rich AGB candidates --- i.e.,
objects with 8 $\mu$m fluxes $<200$ mJy and JHK8 or SAGE colors
consistent with those of the O-rich AGB stars identified here
and in Paper I --- are needed in order to better establish the
relative frequency of O-rich vs.\ C-rich envelopes among the rapidly
mass-losing AGB stars in the LMC.

\buc\ established that four objects that reside within the O-rich AGB
region (\msx\ LMC 412, 1150, 1677, and 1686) are in fact Mira
variables located in the halo of the Milky Way. We have identified
three more such candidate GMVs in the Table 1 sample, based primarily on their
positions in the \tmass-\msx\ color-magnitude diagrams (Fig.\ 4). One
of these objects, \msx\ LMC 1048 (RT Men), was included in a sample of
GMVs studied by \citet{jura92} and was assigned a distance of 4.9
kpc. Another object, \msx\ LMC 362 (ZZ Men), was found by
\citet{wood85} to have a radial velocity inconsistent with membership
in the LMC. The third, \msx\ LMC 716, does not have a variable star
designation but lies in the same region of JHK8
color-magnitude space as the other well-established GMVs. It is
possible that a few additional RSG candidates are in fact GMV stars,
given the overlap in the regions of color-color space occupied by
these classes (over a dozen have ambiguous JHK8 color-based
classifications; Table 1, column 10).

\subsection{C-rich AGB stars} \label{subsec:dis_class_c}

Carbon stars represent the single largest source population in our
sample, with a total of 87 objects identified (69) or tentatively
identified (18) among the 250 Table 1 sources. The total number of carbon
stars may be as high as 93, depending on the nature of the 3 candidate
O-rich AGB stars and the 3 AGB stars whose chemistries are ambiguous
at present. The vast majority of the 88 Table 1 objects originally
classified as OH/IR stars by \egant\ are reclassified here as carbon
stars as a consequence of the application of revised \buc\ IRS
color-color criteria.  As C-rich AGB stars with bluer JHK8 colors tend
to overlap the O-rich AGB stars in color-color space (see Fig.\
\ref{fig:colors}), and AGB stars exhibit considerable photometric
variability, a small percentage of those objects lacking IRS spectra
that are classified here as carbon stars may in fact be O-rich AGB
stars. Furthermore, our cross-check against the available literature
demonstrates that we had misclassified a few O-rich evolved stars as C
stars on the basis of JHK8 colors, so there is the possibility that a
few more such misclassifications remain. Ignoring this slight caveat,
it appears carbon stars constitute at least 26\% and perhaps as much
as $36$\% of the LMC's most luminous 8 $\mu$m sources, where the
former percentage includes only the 65 high-confidence ``C AGB''
classifications.

Intriguingly, several confirmed and candidate carbon
stars (\msx\ LMC 44, 83, 435, 635, 888, 936, and 1130) have
inferred bolometric luminosities in the range
$L_{bol}\sim2.5\times10^4$ $L_\odot$ to $\sim5\times10^4$
$L_\odot$. These stars may belong to the rare but important population
of carbon stars that lie very near the AGB tip (e.g., Kastner et al.\
1993; van Loon et al. 1998b; Frost et al. 1998; van Loon et al.\
1999b). The brightest such object,
\msx\ LMC 888, is particularly noteworthy, as its inferred luminosity
of $\sim5\times10^4$ $L_\odot$ would place it very near the
theoretical AGB luminosity limit ($\sim6\times10^4$ $L_\odot$; Iben \&
Renzini 1983). Recent models predict carbon star formation should be
suppressed at such luminosities as a consequence of ``hot bottom
burning'' (see review in Herwig 2005).  Hence these candidate
high-luminosity carbon stars, and \msx\ LMC 888 in particular, are
worthy of follow-up IR spectroscopy to confirm that they indeed
display C-rich envelope chemistries. Furthermore, the very red colors
of \msx\ LMC 888 and \msx\ LMC 635 (IRAS 05278$-$6942) --- the latter
a carbon star with $L_{bol}\sim4\times10^4$ $L_\odot$ (Groenewegen et
al.\ 2007) and 2MASS/MSX colors redder than those of the carbon stars
surveyed in Paper I --- indicates that more high-$L_{bol}$ carbon
stars may lurk among the very red, unclassified objects in Table 1
(see below).

\subsection{Early-type Emission-line Stars with Dusty Disks}

A number of \msx\ sources appear to be associated with dusty,
early-type, emission-line stars. Several of these IR sources lie in
the same general region of JHK8 color-color space --- i.e., between the RSG
and \HII\ region classification boxes --- as the IR sources
associated with the B[e] supergiants HD 268835 and HD 37974 (Kastner
et al.\ 2006 and references therein).  Like the \msx\
counterparts to these stars (\msx\ LMC 1326 and 890, respectively),
three of these infrared sources --- \msx\ LMC 262, 323, and 1438 ---
are associated with A, B, or B[e] supergiants.  A handful of
additional \msx\ sources in this general region of color-color space
--- i.e., \msx\ LMC 134, 198, 646, and 1296 --- also appear to be
associated with optically luminous, early-type, and/or emission-line
stars (Table 1). Another source in this color-color region, \msx\ LMC
344 (HD 35231), was tentatively identified as an O-rich evolved star
by van Loon et al.\ (2005a), but its JHK8 colors and its association
with a visually bright (HD catalog) star suggest it may also be a
dusty early-type emission-line object (we leave it unclassified at
present). Meanwhile four other stars (\msx\ LMC 224, 461, 585, and
773) have the JHK8 colors of RSGs, but all have SIMBAD associations
with early-type stars; \msx\ LMC 224 and 773 are associated with
Wolf-Rayet [WR] stars.

Some or perhaps all of these objects could be luminous, early-type
stars encircled by dusty disks. Provided the positional associations
of any or all of the dozen or so Table 1 mid-IR sources that lie near
early-type, emission-line stars can be confirmed, the ``dusty disk''
hypothesis is testable via mid-IR spectroscopy combined with modeling
of the stars' infrared spectral energy distributions (e.g., Kastner et
al.\ 2006). Note that we classify \msx\ LMC 198, 461, 646, and 1296 as
candidate \HII\ regions at present, due to their faint near-IR
magnitudes (LMC B[e] supergiants and WR stars typically display JHK
magnitudes $\stackrel{<}{\sim}10$), ambiguity in their SIMBAD spectral
types, and/or a high probability of misidentification due to crowded
fields.

\subsection{Unclassified Objects} \label{subsec:dis_unclass}

Out of our sample of 250 objects, only 10 ($\sim4$\% of the sample)
neither fall into any of the Table 2 color-color regions nor have been
previously classified in the literature. These objects --- most of
which are found above the \HII\ region boxes and/or to the right of
the O-AGB, C-AGB, RSG, and GMV boxes in JHK8 color-color space
(Fig.~\ref{fig:colorJHKA}) --- therefore remain unclassified. Two of
these objects (\msx\ LMC 466 and 1379) were classified as candidate
OH/IR stars by \egant. Given that we find most such candidate OH/IR
stars are in fact C-rich AGB stars (\S 4.3), these two
sources are most likely carbon stars with very high mass-loss
rates. All of these unclassified objects are worthy of IR spectroscopy
aimed at ascertaining their natures.

\section{SPITZER IRAC/MIPS COLORS OF THE SAMPLE OBJECTS}

\begin{figure}
\epsscale{1.05}
\plotone{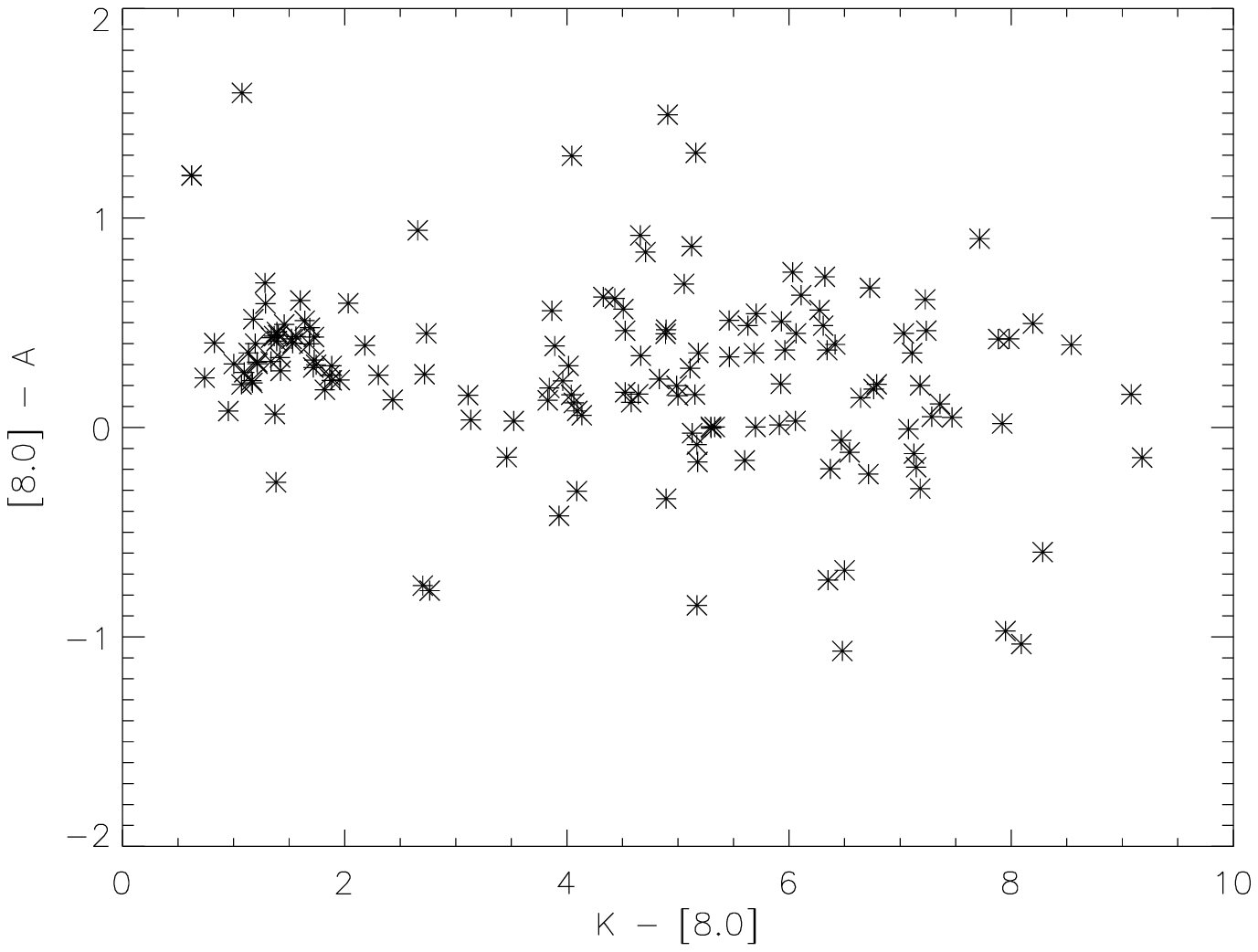}
\caption{Comparison of \msx\ A-band and {\it Spitzer} IRAC 8.0 $\mu$m
  magnitudes for the Table 1 stars for which SAGE PSC photometry is
  available (Table 4). The ordinate is $K-[8.0]$ color and the abscissa is the
  difference between MSX A and IRAC 8.0 $\mu$m magnitudes. Note the
  tight cluster of RSGs centered at $K-[8.0] \sim 1.5$ and the looser
  distribution of objects with $K-[8.0] > 2$. The latter group is
  dominated by AGB stars, which display larger amplitudes of variation
  than the RSGs. See text.}
\label{fig:A8comp}
\end{figure}

In \buc, we demonstrated how the IRS spectral classifications of
luminous LMC mid-IR sources could be used to classify sources on the
basis of {\it Spitzer} IRAC/MIPS (as well as \tmass/IRAC/MIPS)
color-color diagrams. To explore this potential, we used the SAGE IRAC
and MIPS point source catalogs\footnote{The SAGE point source catalogs
  are available via the GATOR interface to the NASA/IPAC Infrared
  Science Archive
  (http://irsa.ipac.caltech.edu/applications/Gator/). } to identify
counterparts to Table 1 sources. Specifically, we selected all SAGE
point sources with IRAC 8.0 $\mu$m magnitudes $\le 6.5$ that have MIPS
24 $\mu$m counterparts, and then cross-correlated their positions with
the \tmass\ positions of the Table 1 sample, so as to identify those
SAGE sources within $7''$ (i.e., the approximate MIPS 24 $\mu$m PSF
FWHM) of a Table 1 object. This procedure resulted in a subsample of
172 SAGE counterparts to Table 1 sources (Table 4).  Most of the Table
1 objects not present in Table 4 are \HII\ regions that were
rejected from the SAGE PSC as a consequence of their spatial
resolution by IRAC (\S 3.2.1). Source variability likely accounts for
most of the remaining omissions of Table 1 sources from the SAGE
subsample.  A plot of the difference between MSX A-band magnitude and
IRAC 8.0 $\mu$m magnitude vs.\ $K-[8.0]$ color (Fig.~\ref{fig:A8comp})
illustrates the degree of 8 $\mu$m variability exhibited by the Table
4 sources. Figure~\ref{fig:A8comp} also shows an apparent systematic
$\sim0.3-0.4$ mag discrepancy between \msx\ and IRAC 8 $\mu$m
magnitudes. This discrepancy is likely a result of the systematically
larger A-band fluxes in the version 6 \msx\ data (\S 2).

In Figure~\ref{fig:JHK8SAGE} we plot the positions of the SAGE
counterparts to Table 1 sources on JHK8 color-color diagrams
constructed from \tmass\ J, H, K magnitudes and IRAC 8.0 $\mu$m
magnitudes. This Figure demonstrates that the
\tmass+\msx\ JHK8 source classification regions (Table 3)
are directly applicable to the combination of \tmass\ (J, H, K) and IRAC
(8.0 $\mu$m) data. The only significant modification to the Table 3 JHK8
classification criteria that is suggested by the
Figure is a ``blueward'' increase of the area bounding O-rich AGB
stars by $\sim1.0$ mag in K--[8.0], so as to encompass three such
sources that lie outside of the ``O AGB'' region.

In Figs.~\ref{fig:SAGEcolcol}, \ref{fig:IRACcolcol}, and
\ref{fig:2MASSIRAC} we display additional color-color diagrams that make
use of {\it Spitzer} IRAC/MIPS data, IRAC data only, and \tmass/IRAC
data, respectively. In each of these diagrams, carbon stars and RSGs
appear as distinct clusters, verifying the \buc\ assertion that
\tmass-{\it Spitzer} color-color diagrams can be used to identify
candidate C-rich AGB stars and RSGs among samples of luminous mid-IR
point sources detected by IRAC/MIPS. Furthermore --- in contrast to
the extended region of overlap between C-rich and O-rich AGB stars
seen in the JHK8 diagrams --- those AGB stars whose O-rich nature has
been firmly established via IRS spectroscopy appear as tight
groupings, distinct from the C-rich AGBs, in both the IRAC/MIPS
(Fig.~\ref{fig:SAGEcolcol}) and IRAC-only (Fig.~\ref{fig:IRACcolcol})
color-color diagrams. A similar degree of separation between C-rich
and O-rich objects was apparent in IRAC/MIPS color-color diagrams
presented in \buc\ and Lagadec et al.\ (2007), but the diagrams
presented here include many more objects than the former, while the
sample considered in the latter paper did not include RGSs and was not
subject to a uniform minimum 8 $\mu$m luminosity criterion, as is the
case here. 
Given the particularly sharp separation between C-rich and
O-rich objects in Figure~\ref{fig:SAGEcolcol}, we present in Table 5 a
set of IRAC/MIPS color-color classification criteria for C-rich AGB,
O-rich AGB, and RSG stars; the corresponding source classification
regions are indicated in Fig.~\ref{fig:SAGEcolcol}. 

{\it Spitzer} color-color diagrams such as those in
Figures~\ref{fig:SAGEcolcol} and ~\ref{fig:IRACcolcol} may be of more limited
use in identifying the more extreme and/or exotic evolved stars among
samples of luminous mid-IR sources. Such objects (e.g., dusty B[e]
stars) mainly ``contaminate'' the C-rich AGB group. However, the B[e]
star IR sources do form a distinct grouping in the the \tmass/IRAC
color-color diagram displayed in Figure~\ref{fig:2MASSIRAC}. This
color-color diagram furthermore is of particular interest due to its
potential broad utility in the {\it Spitzer} ``warm mission'' era,
during which only the first two IRAC channels at 3.6 and 4.5 $\mu$m
will generate scientitically useful image data. Indeed, comparison of
Figs.~\ref{fig:JHK8SAGE} and ~\ref{fig:2MASSIRAC} demonstrates that
the combination of \tmass and ``warm'' {\it Spitzer} photometry at 3.6
and 4.5 $\mu$m will provide source classification capabilities similar
to those provided by the JHK8 system.

\section{SUMMARY AND CONCLUSIONS} \label{sec:con}

To ascertain the nature of the most luminous mid-infrared sources in
the LMC, we have applied a revised version of the Buchanan et al.\
(2006) (\buc) infrared (JHK8) color classification scheme --- which is
based on the results of Spitzer IRS spectroscopy of a representative
sample of $\sim60$ objects selected from among the catalog of
$\sim1650$ \tmass/\msx\ sources compiled by \egant\ --- to all 250
\egant\ LMC sources satisfying the \buc\ 8 $\mu$m flux limit ($F_{8.3}
\stackrel{>}{\sim} 200$~mJy; A $\le$ 6.5) for which \tmass\ fluxes (or
flux upper limits) are available. We have augmented, cross-checked,
and revised (where necessary) the resulting JHK8 color-based
classifications via a number of independent means (\S 3.2.1).

We thereby obtain the following results: 51 sources are confidently
identified as evolved stars with oxygen-rich envelopes, where 42 of
these are red supergiants and 9 are O-rich AGB stars; 69 sources are
confidently identified as C-rich AGB stars; 58 sources are confirmed
as H {\sc ii} regions; one source is a planetary nebula; 4 sources are
dusty B[e] supergiants; and 7 objects are foreground Mira variables in
the halo of the Milky Way. Another 47 objects are tentatively
classified as either C-rich AGB (18), O-rich AGB (3), RSGs (2), H {\sc
  ii} regions (19), or dusty, early-type, emission-line stars (B[e]
supergiants or Wolf-Rayet stars; 6 objects). Including these tentative
classifications, we find the following proportions of objects among
the LMC's most 243 most luminous compact 8 $\mu$m sources (not
including the 7 foreground, Milky Way AGB stars): 35\% (87 objects)
are C-rich AGB stars, 5\% (12 objects) are O-rich AGB stars, 18\% (44
objects) are red supergiants, 31\% (77 objects) are compact H {\sc ii}
regions, and 4\% (10 objects) are associated with early-type,
emission-line stars (including one planetary nebula). Three additional
sources are also most likely LMC AGB stars, but their circumstellar
chemistries are unknown. Only 10 objects ($4$\%) cannot be classified
or tentatively classified based on their positions in \tmass-\msx\
color-color/color-magnitude diagrams and/or other means.  Comparison
of the classifications obtained via the JHK8 color-color and
color-magnitude criteria (column 10 of Table 1) with classifications
obtained via independent means (\S 3.2.1) indicates the ``success
rate'' of the JHK8 classification system is at least $70$\%.

The large ratio of C-rich to O-rich AGB stars ($\sim7:1$, including
confirmed and candidate objects of both classes) confirms that
carbon stars are common, while O-rich AGB stars are quite rare, among
the most luminous 8 $\mu$m LMC sources. This result is consistent with
the hypothesis that carbon stars form easily in low-metallicity
environments, due to the relative ease of inverting the C/O ratio at
the stellar surface (see review by Herwig 2005). The relatively
small ratio of AGB stars to RSGs among our sample ($\sim2.3:1$, including
confirmed and candidate AGB stars and RSGs) is in
keeping with the expectation that, in selecting the most luminous 8
$\mu$m sources among the population of highly evolved stars in the
LMC, we are sampling the intermediate-to-high range of initial stellar
masses.

We applied ``bolometric corrections,'' established
empirically in \buc, to estimate the integrated infrared
luminosities of carbon stars and RSGs based on their K-band
magnitudes. Because our sample includes all luminous
8~\micron\ LMC sources with \tmass\ counterparts, these
estimates directly yield the peak IR luminosities reached by
relatively ``blue'' RSGs (i.e., those with $K-A \stackrel{<}{\sim}3$)
and those LMC carbon stars which have $K-A
\stackrel{<}{\sim}6$. The most luminous such RSGs in the LMC
therefore appear to be \msx\ LMC 1679 and 43, with 
estimated luminosities of $2.5\times10^5$ $L_\odot$ and
$1.9\times10^5$ $L_\odot$ respectively. These
luminosities fall well short of that of the OH/IR supergiant
\msx\ LMC 1182, with $L_\star \sim
5\times10^5$ $L_\odot$. It remains possible
that a few more very luminous and highly obscured red supergiants
lurk among the unclassified and/or tentatively classified objects in our
sample. Meanwhile, the most luminous carbon stars in our
sample rise very high on the AGB; the C-rich AGB candidate
\msx\ LMC 888 ($L_\star \sim
5\times10^4$ $L_\odot$) approaches theoretical expectations for the
peak luminosities of carbon stars (Herwig 2005).

Examination of {\it Spitzer} IRAC/MIPS (SAGE) photometry available for
the Table 1 sources demonstrates that IRAC/MIPS color-color diagrams,
augmented by near-IR survey data, should greatly facilitate the
classification of the large number of luminous mid-IR point sources
that have been and will be detected via {\it Spitzer} imaging of the
Milky Way and external galaxies. In particular, we find that both
IRAC-only ($[3.6]-[4.5]$ vs.\ $[5.0]-[8.0]$) and IRAC/MIPS
($[5.8]-[8.0]$ vs.\ $[8.0]-[24]$) color-color diagrams provide 
effective means to identify and distinguish between C-rich and O-rich
evolved stars. These results argue for the broad application of such
{\it Spitzer} IRS-based IR color-color classification techniques to
other Local Group galaxies whose most luminous IR point sources are
detectable and resolvable by \tmass\ and/or {\it Spitzer}. In this
way, one might determine and compare the number distributions of
mid-IR-luminous sources spanning a wide range of metallicity and star
formation history, so as to understand the dependence of the
populations of various classes of rapidly mass-losing evolved stars,
as well as young and/or dusty massive stars, on galactic environment
and star formation rate. Such work can continue unabated during the
{\it Spitzer} ``warm mission,'' via the combination of
ground-based near-IR and short-wavelength IRAC photometry.

Further IR spectroscopic observations of the
unclassified and tentatively classified sources among the sample of
luminous 8~\micron\ LMC sources studied here are required to determine
unambiguously the natures of these sources. Meanwhile, IR
spectroscopic observations of objects with \tmass/\msx colors similar
to those of the few O-rich AGB stars in Table 1, but with smaller 8
$\mu$m luminosities, would establish whether, and to what degree, our
selection criteria bias our finding of a very large ratio of C-rich to
O-rich AGB stars in the LMC.

\acknowledgments{The authors thank Bill Forrest and the referee, Jacco
  van Loon, for their extensive and incisive
comments on this manuscript. Support for this research was
provided by JPL/Caltech Spitzer Space Telescope General
Observer grant NMO710076/1265276 to R.I.T. and NMO710076/1264276 to
Valparaiso University.} 

{\it Facilities:} \facility{\spitzer}


\begin{figure}
\epsscale{1.0}
\plotone{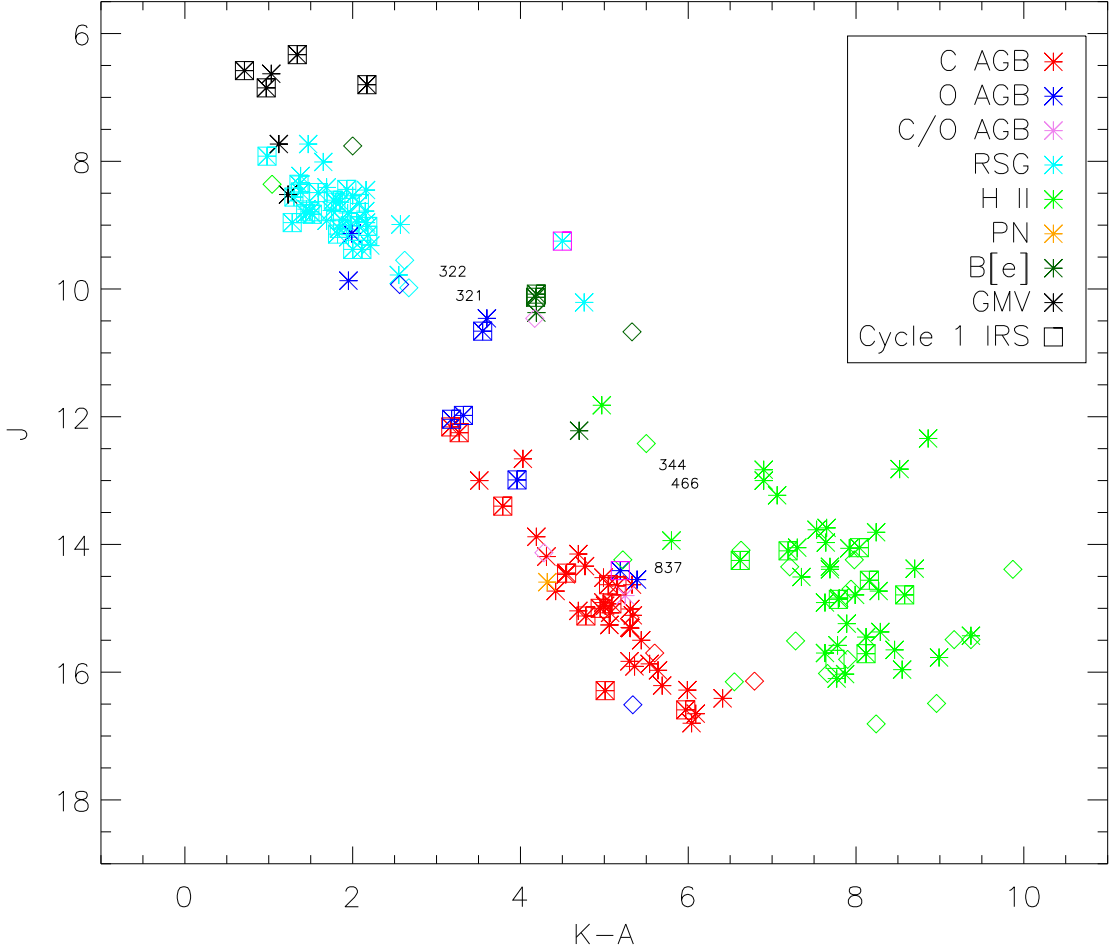}
\caption{{\it (a)} \protect\tmass/\protect\msx\ $J$
  vs. $K-A$ color-magnitude diagram for the Table 1 sample. 
  Objects with $J$ band upper limits have been omitted. The symbol
  color coding is based on classifications in column 11 of Table 1.
  Objects classified with high confidence are indicated by
  asterisks; objects with tentative classifications are indicated by
  diamonds. Points corresponding to objects included in the Paper I IRS
  sample are enclosed in squares. Objects with no
  classification in column 11 of Table 1 are indicated by MSX number.}
\label{fig:magcolors}
\end{figure}

\begin{figure}
\epsscale{1.0}
\addtocounter{figure}{-1}
\plotone{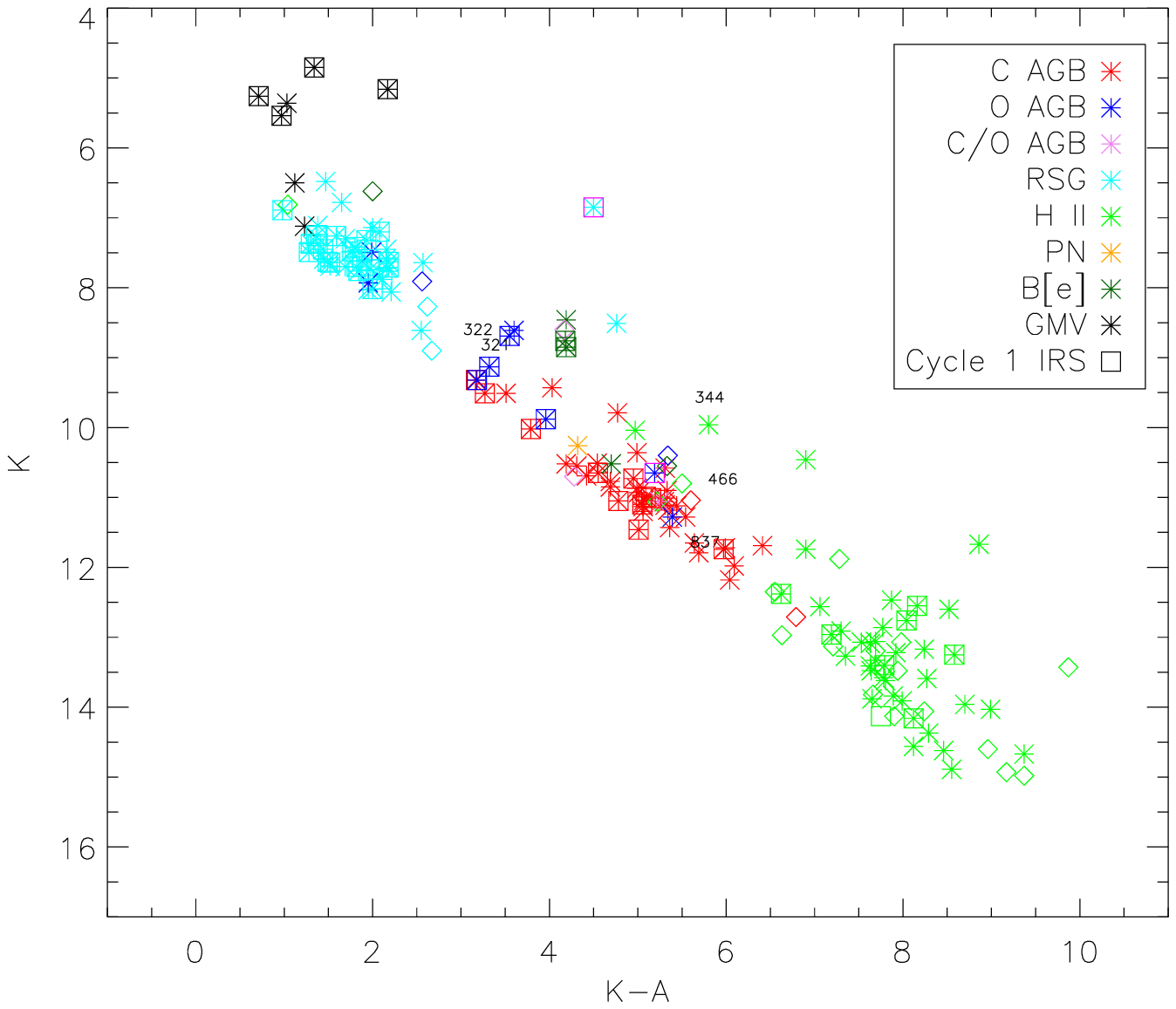}
\caption{{\it (b)} As in Figure~\ref{fig:magcolors} {\it (a)}, but for
  $K$ vs. $K-A$, omitting objects with $K$ band upper limits.}
\end{figure}

\begin{figure}
\includegraphics[scale=1.0,angle=90]{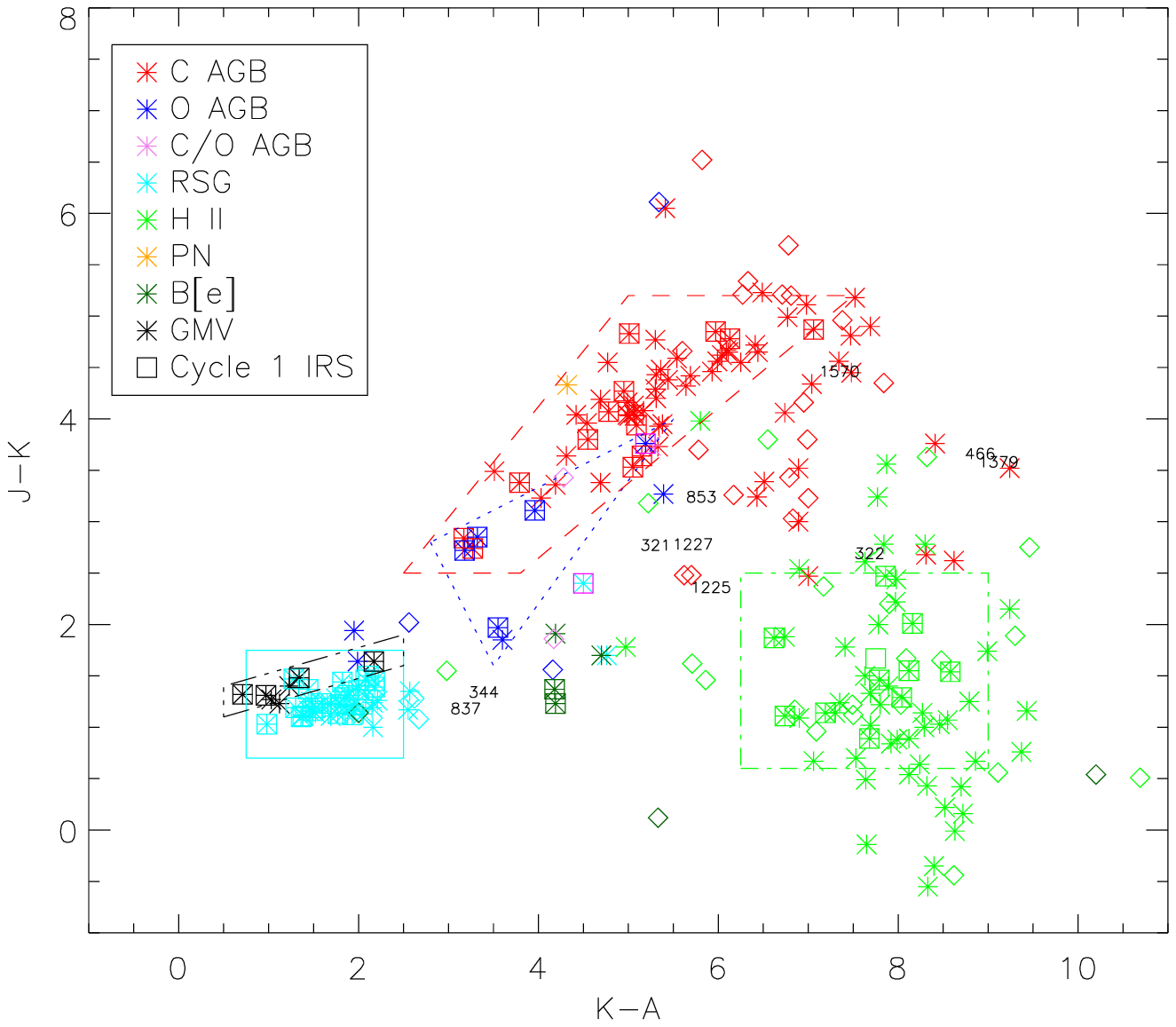}
\caption{{\it (a)} $J-K$ vs.\ $K-A$ color-color diagram for the Table
  1 sample, color-coded to illustrate the results of classification
  based on the \tmass/\msx\ color-color and color-magnitude diagrams
  combined with a cross-check of the available literature (column 11
  of Table 1). Revised \buc\ color-color classification
  regions (\S 3.1) are
  overlaid. Symbol meanings are as in Figure~\ref{fig:magcolors}.
  }
\label{fig:colorJHKA}
\end{figure}

\begin{figure}
\addtocounter{figure}{-1}
\includegraphics[scale=1.0,angle=90]{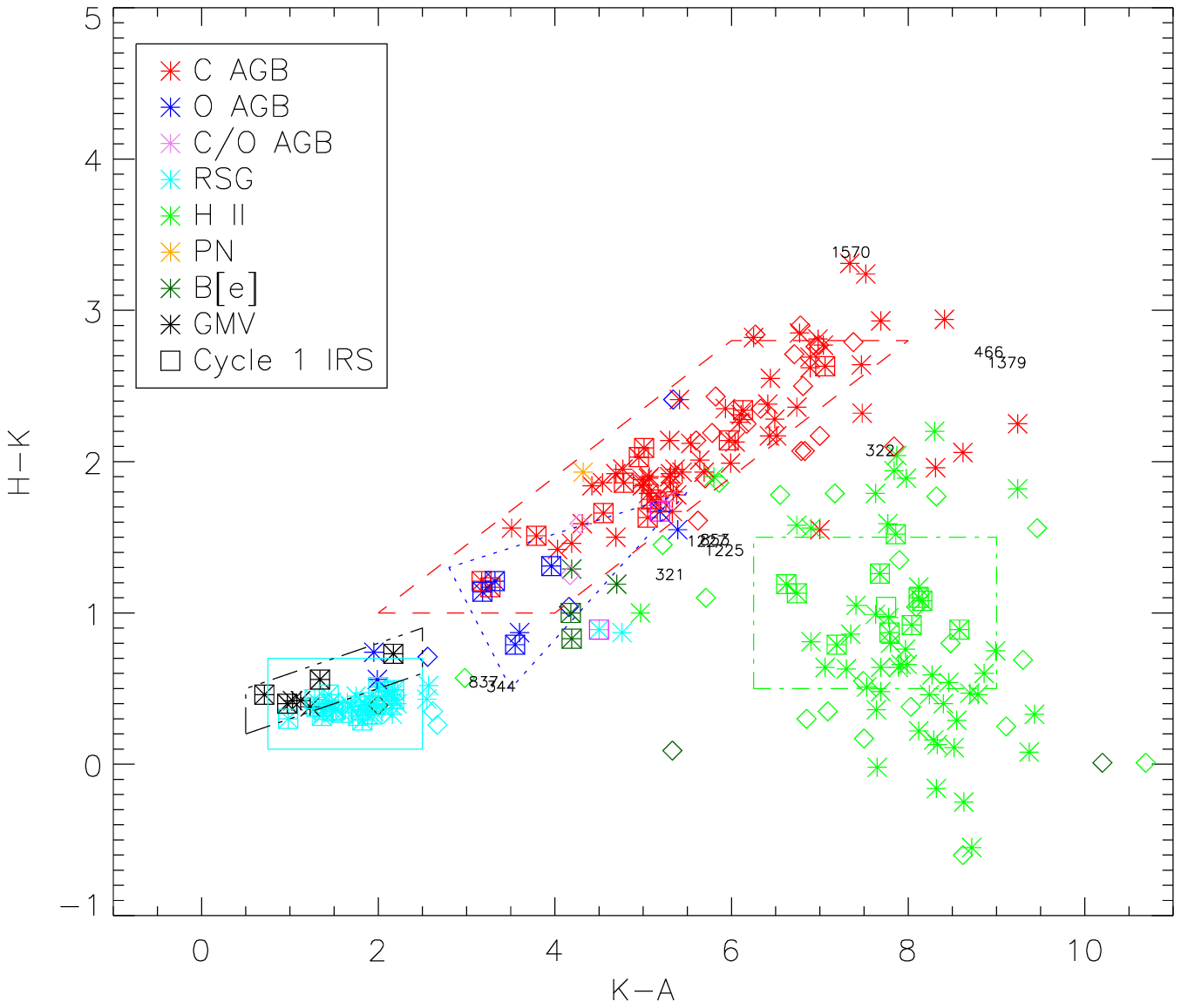}
\caption{{\it (b)} As in Figure~\ref{fig:colorJHKA} {\it (a)}, but for
  $H-K$ vs. $K-A$ colors.}  
\end{figure}

\begin{figure}
\epsscale{0.8}
\plotone{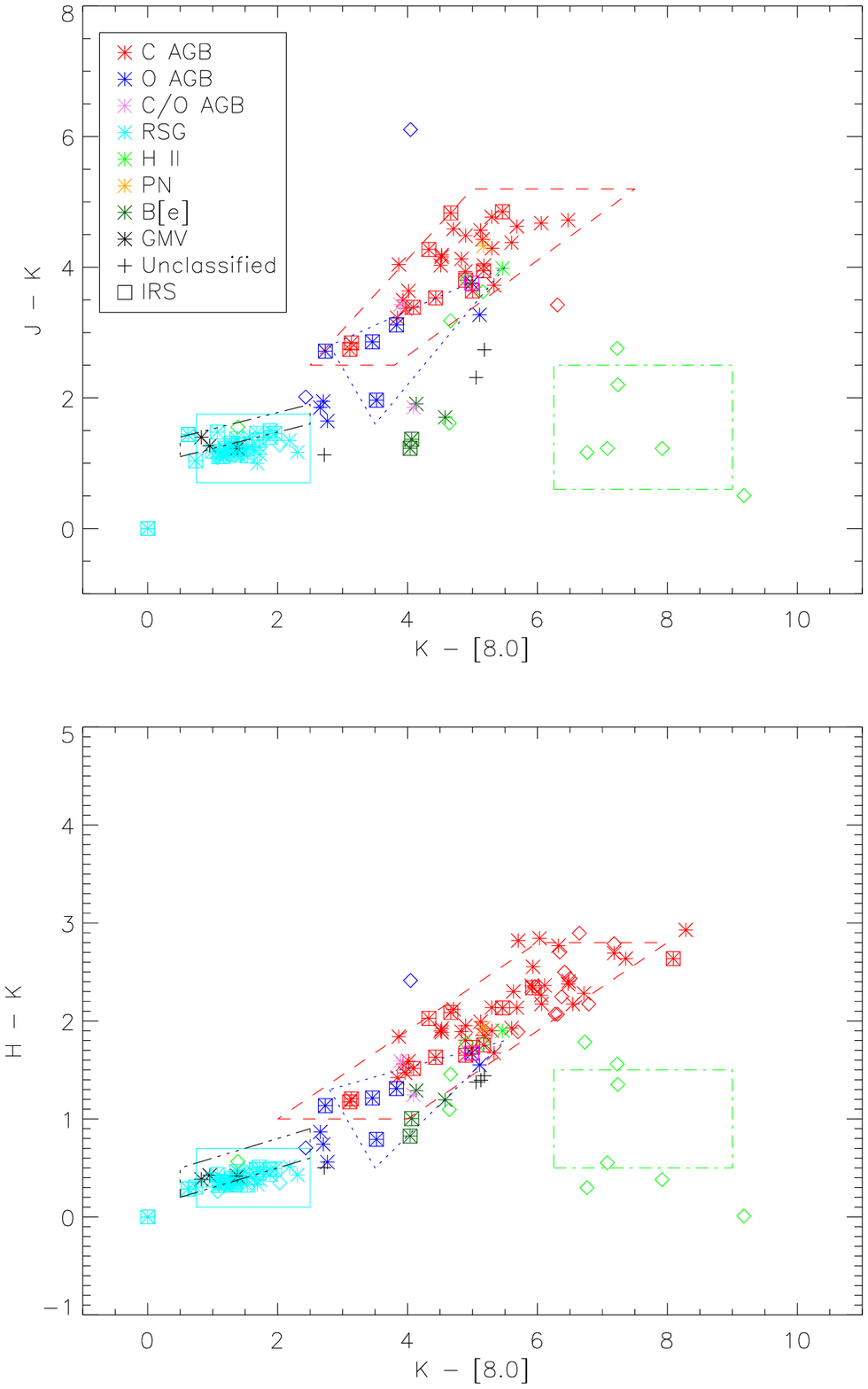}
\caption{JHK8 color-color diagrams for the Table 1 objects as
  constructed from available \tmass\ and {\it Spitzer} (SAGE) data
  (Table 4), overlaid with \tmass-\msx-based object classification
  regions (Table 2). Symbol meanings are as in
  Figure~\ref{fig:magcolors}.  Top: $J-K$ vs.\ $K-[8.0]$; bottom:
  $H-K$ vs.\ $K-[8.0]$. Objects for which \tmass\ data are upper
  limits are not plotted.}
\label{fig:JHK8SAGE}
\end{figure}

\begin{figure}
\epsscale{1.05}
\plotone{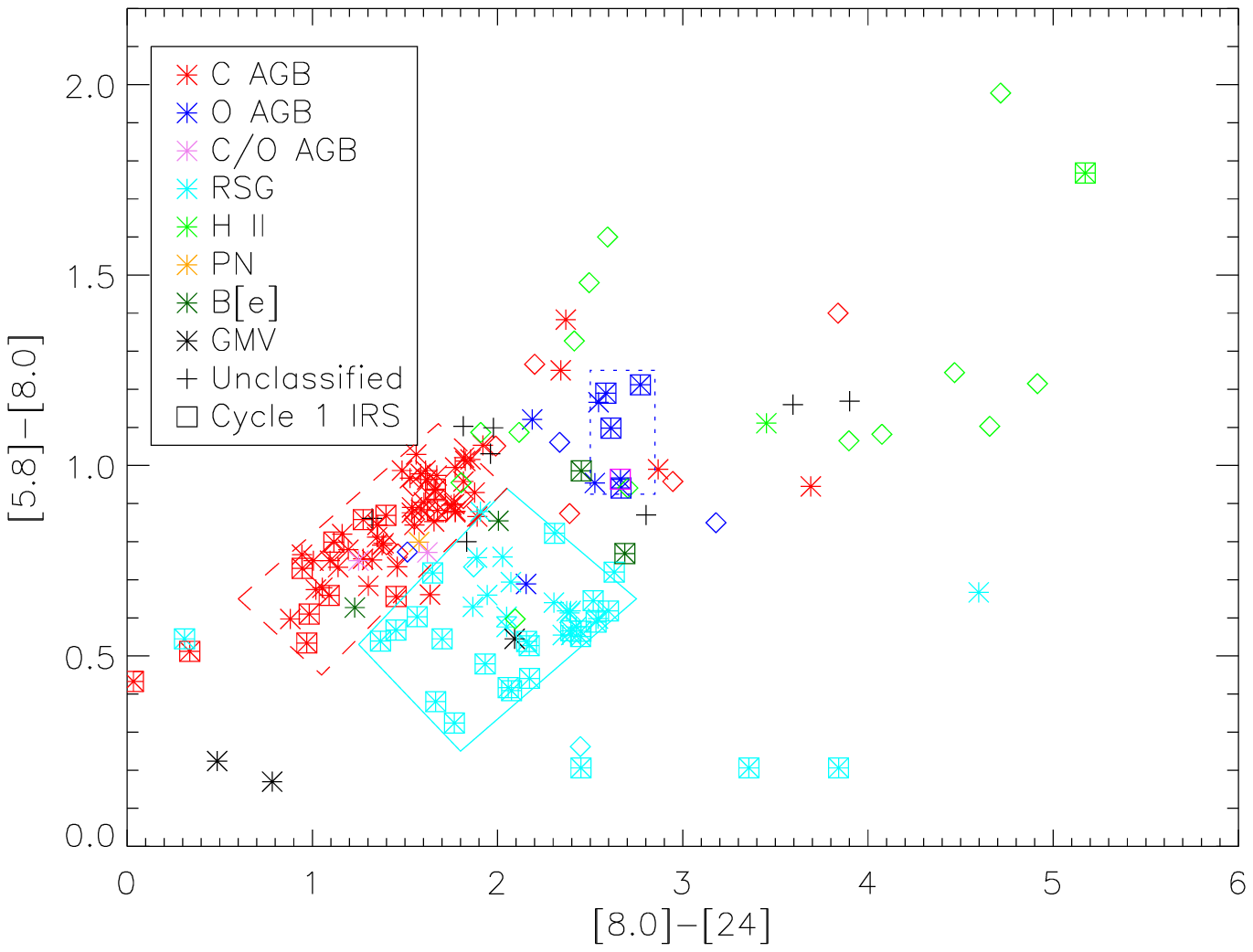}
\caption{{\it Spitzer} IRAC/MIPS $[5.8]-[8.0]$ vs.\ $[8.0]-[24]$
  color-color diagram for the Table 4 objects. Symbol meanings are as
  in Figure~\ref{fig:magcolors}. Boxes indicate the Table 5 IRAC/MIPS
  color criteria for classifying IR-luminous RSGs, C-rich AGB stars,
  and O-rich AGB stars in the LMC. The three red supergiants near the
  bottom of the plot evidently have unreliable IRAC fluxes, perhaps as
  a result of detector saturation. }
\label{fig:SAGEcolcol}
\end{figure}

\begin{figure}
\epsscale{1.05}
\plotone{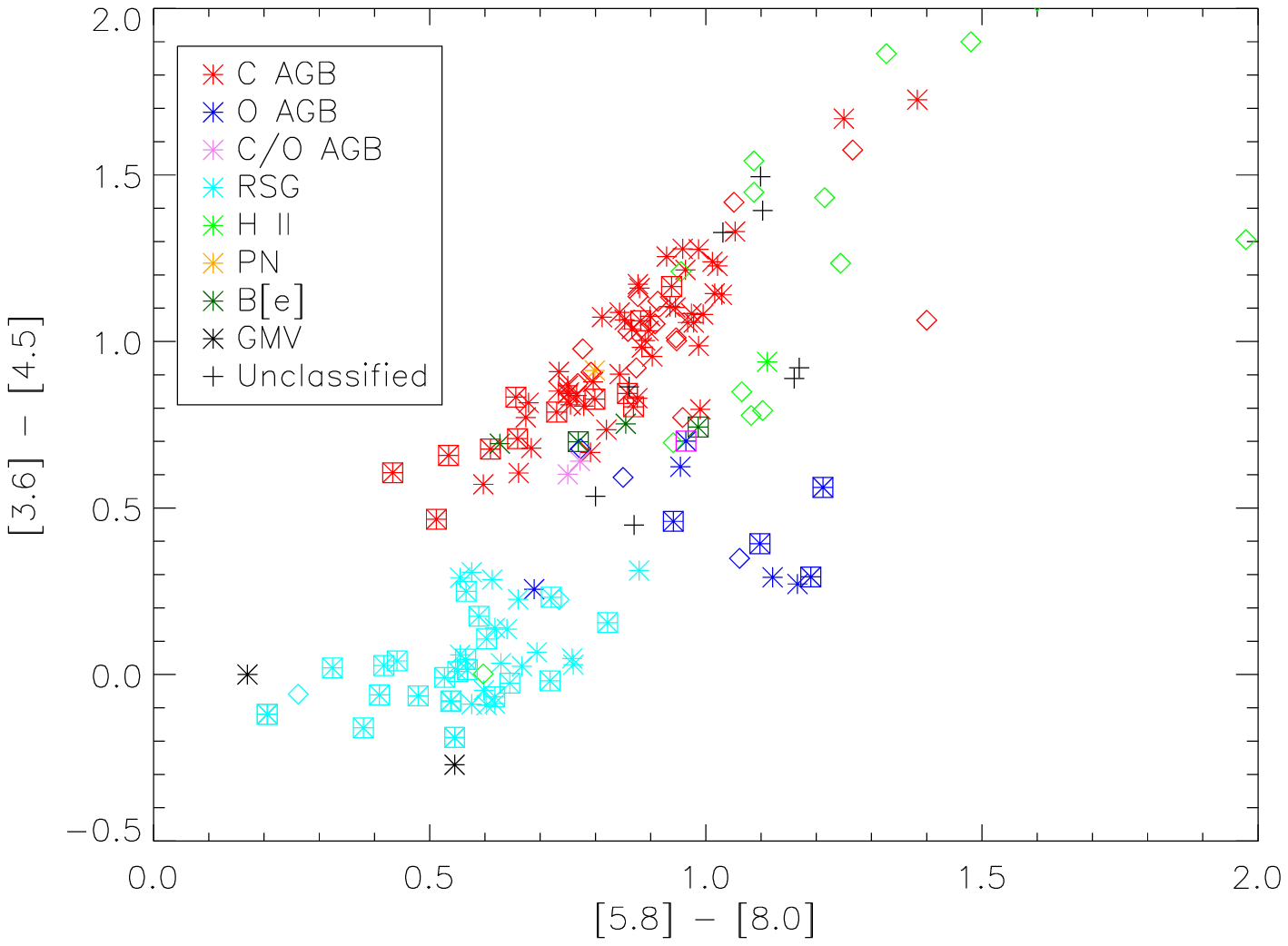}
\caption{{\it Spitzer} IRAC $[3.6]-[4.5]$ vs.\ $[5.8]-[8.0]$
  color-color diagram for the Table 4 objects. Symbol meanings are as in
  Figure~\ref{fig:magcolors}. }
\label{fig:IRACcolcol}
\end{figure}

\begin{figure}
\epsscale{1.05}
\plotone{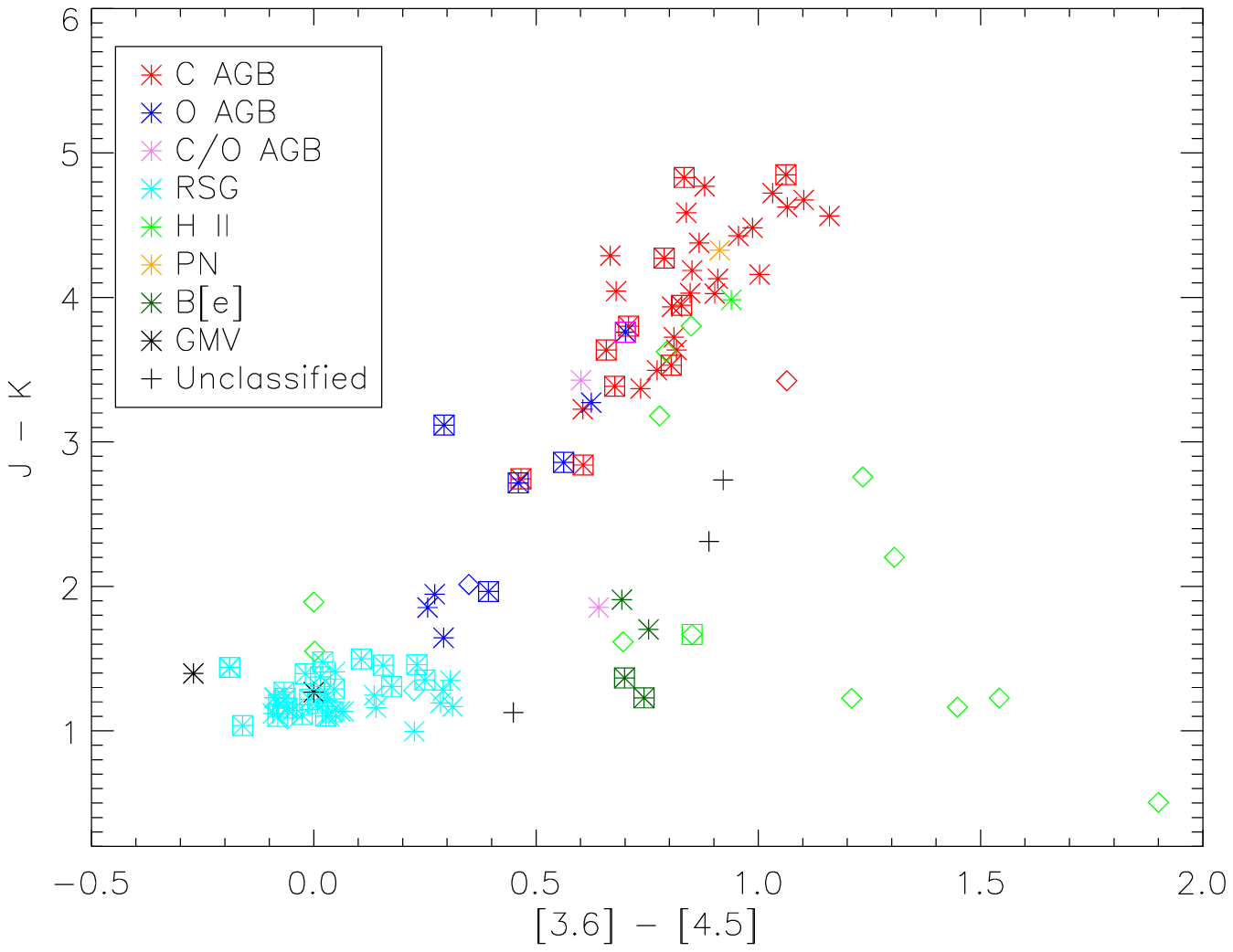}
\caption{\tmass\ J$-$K vs.\ {\it Spitzer} IRAC $[3.6]-[4.5]$ 
  color-color diagram for the Table 4 objects. Symbol meanings are as in
  Figure~\ref{fig:magcolors}. }
\label{fig:2MASSIRAC}
\end{figure}


 commands

\tablerefs{1. Matsuura et al. 2005; 2. van Loon et al. 2006; 3. SIMBAD; 4. Oliveira et al. 2006; 
5. {\protect\buc}; 6. Zijlstra et al. 2006; 7. de Winter et al. 2001; 
8. Indebetouw et al. 2004; 9. Sanduleak et al. 1978; 
10. van Loon et al. 2001b; 11. Westerlund et al. 1981; 12. Kastner et al. 2006;
13. van Loon et al. 2005a; 14. van Loon et al. 1998}


\tablenotetext{a}{Source types and names obtained from SIMBAD
(simbad.u-strasbg.fr/sim-fid.pl).  Object types (other than stellar spectral types)
are as follows: V* = variable
star; EmO = emission-line source; IR = infrared source,
* = star, sr* = semi-regular variable star; Mi* = variable star of
Mira type; OpC = open cluster; As* =
association of stars, Cl* = cluster of stars; *iC =
star in cluster; *i* = star in double
system; MoC = molecular cloud; Rad = radio
source; *iA = star in association; HV* =
high-velocity star. } 

\tablenotetext{b}{\tmass\ source magnitudes followed by a `u' designate flux upper limits.}

\tablenotetext{c}{JHK8 color-based classifications (see \S\S 3.1,
  3.2). Colons indicate tentative classifications; blank entries
  indicate sources that are not classifiable via JHK8 colors.}

\tablenotetext{d}{Most probable classifications for sources, based on
  JHK8 color-based classification system, available literature, and additional
  cross-checks (see \S 3.2). Question marks indicate tentative classifications;
  blank entries indicate sources that have no classifications. } 

\tablenotetext{e}{{\lir} determined in {\buc}.}

\tablenotetext{f}{Source appears in the expanded {\HII} classification box.}

\tablenotetext{g}{K magnitude was used to determine classification as
  ``RSG'' or ``GMV'' (see \S 3.2.1).}

\tablenotetext{h}{{\lir} uncertain because K magnitude is an upper limit.}

\tablenotetext{i}{Archival {\it Spitzer} IRS spectrum available (see \S 3.2.1).}



\end{deluxetable}

\begin{deluxetable}{cccc}
\tabletypesize{\small}
\tablewidth{0pt}
\tablecolumns{4}
\tablecaption{Summary of classifications of luminous 8
  $\mu$m sources in the LMC\label{tab:summarytable}}
\tablehead{
\colhead{Class} &
\colhead{EVP01} &
\colhead{\buc} &
\colhead{this paper\tablenotemark{a}} \\
}
\startdata
All           &  250\tablenotemark{b}  &  57                  &  250 \\
C AGB         &  8\tablenotemark{c}    &  13                  &  87 (18) \\
O AGB         &  21                    &  4                   &  12 (3) \\
C/O AGB\tablenotemark{d} &  \nodata               &  \nodata             &  3 (2) \\
RSG           &  19                    &  21                  &  44 (2) \\
\HII          &  39                    &  11                  &  77 (19) \\
GMV           &  0                     &  4                   &  7 \\
PN            &  34                    &  0                   &  1\tablenotemark{e} \\
OH/IR         &  88                    &  2                   &  2\tablenotemark{f} \\
B[e]          &  0                     &  2\tablenotemark{g}  &  7 (3) \\
WR            &  0                     &  0                   &  (2) \\
Unclassified  &  41                    &  0                   &  10  \\

\enddata 

\tablenotetext{a}{Numbers outside parentheses indicate total number of
  confirmed plus tentative classifications. Numbers within parentheses
  indicate number of tentative classifications.}

\tablenotetext{b}{Only includes objects considered in this paper.}

\tablenotetext{c}{Includes objects classified as either ``C AGB'' or
  ``C IR''.} 

\tablenotetext{d}{Objects that fall in the zone of confusion between
  O-rich and C-rich AGB star JHK8 color classification regions.}

\tablenotetext{e}{A small
  fraction of objects classified as \HII\ regions may be
  PNs; see \S 4.1.} 

\tablenotetext{f}{One AGB star (included in ``O AGB'' total) and one RSG (included in ``RSG'' total).}  

\tablenotetext{g}{Two B[e] supergiants in \buc\ were classified as Peculiar.}


\end{deluxetable}

\begin{deluxetable}{lcc}
\rotate
\tabletypesize{\scriptsize}
\tablewidth{0pt}
\tablecolumns{3}
\tablecaption{Luminous LMC Mid-IR Sources: Revised JHK8 Color Classification Criteria}
\tablehead{
\colhead{Class} &
\multicolumn{2}{c}{Criteria}  \\
}

\startdata
\multicolumn{3}{c}{J$-$K vs. K$-$A colors}  \\
RSG              &  0.75 $\le$ (K$-$A) $\le$ 2.5                &  0.7 $\le$ (J$-$K) $\le$ 1.75  \\
O AGB            &  [$-$0.7 $\times$ (J$-$K) + 4.69] $\le$ (K$-$A) $\le$ [0.87 $\times$ (J$-$K) +2.02] &  (J$-$K) $\le$ [0.44 $\times$ (K$-$A) + 1.56]  \\
C AGB            & [0.93 $\times$ (J$-$K) + 0.175] $\le$ (K$-$A) $\le$ [1.37 $\times$ (J$-$K) + 0.375] &  2.5 $\le$ (J$-$K) $\le$ 5.2 \\
\protect\HII     &  6.25 $\le$ (K$-$A) $\le$ 9.0                &  0.6 $\le$ (J$-$K) $\le$ 2.5  \\
Expanded {\HII}  &  6.0 $\le$ (K$-$A) $\le$ 9.5                 &  (J$-$K) $\le$ 3.0  \\
GMV              &  0.5 $\le$ (K$-$A) $\le$ 2.5                 & [0.25 $\times$ (K$-$A) + 0.975] $\le$ (J$-$K) $\le$ [0.25 $\times$ (K$-$A) + 1.275] \\
\\
\multicolumn{3}{c}{H$-$K vs. K$-$A colors}  \\
RSG              &  0.75 $\le$ (K$-$A) $\le$ 2.5                &  0.1 $\le$ (H$-$K) $\le$ 0.7  \\
O AGB            & [$-$1.00 $\times$ (H$-$K) + 4.10] $\le$ (K$-$A) $\le$ [0.91 $\times$ (H$-$K) + 2.95] &  (H$-$K) $\le$ [0.19 $\times$ (K$-$A) + 0.78] \\
C AGB            & [2.22 $\times$ (H$-$K) $-$ 0.22] $\le$ (K$-$A) $\le$ [2.22 $\times$ (H$-$K) + 1.78] &  1.0 $\le$ (H$-$K) $\le$ 2.8 \\
\protect\HII     &  0.75 $\le$ (K$-$A) $\le$ 2.5                &  0.5 $\le$ (H$-$K) $\le$ 1.5  \\ 
Expanded {\HII}  &  6.0 $\le$ (K$-$A) $\le$ 9.5                 &  (H$-$K) $\le$ 2.0   \\
GMV              &  0.5 $\le$ (K$-$A) $\le$ 2.5                 &  [0.2 $\times$ (K$-$A) + 0.1] $\le$ (H$-$K) $\le$ [0.2 $\times$ (K$-$A) + 0.4]  \\
\enddata 

\end{deluxetable}

\begin{deluxetable}{cccrrrrr}

\tablecolumns{8}

\tabletypesize{\footnotesize}

\tablecaption{SAGE IRAC/MIPS Counterparts to Luminous LMC 8 $\mu$m MSX Sources}

\tablehead{ & & & \multicolumn{5}{c}{SAGE (IRAC/MIPS) magnitudes} \\
\colhead{MSX LMC} & 
\colhead{Class\tablenotemark{a}} &
\colhead{SAGE ID} &
\colhead{[3.6]} &
\colhead{[4.5]} &
\colhead{[5.8]} &
\colhead{[8.0]} &
\colhead{[24]} 
}

\startdata
   7   &   RSG  &  J050943.57-652159.3 &  7.39 &  7.33 &  7.04 &  6.48 &  4.13 \\
  43   &   RSG  &  J050414.12-671614.4 &  6.20 &  6.29 &  6.02 &  5.45 &  3.40 \\
  44   & C AGB  &  J051110.46-675210.6 &  8.08 &  7.05 &  6.11 &  5.22 &  3.47 \\
  45   & C AGB  &  J051041.21-683606.6 &  9.26 &  8.04 &  7.00 &  6.03 &  4.41 \\
  46   &  HII?  &  J050354.56-671848.5 & 10.12 &  9.27 &  8.51 &  7.44 &  3.55 \\
  47   & C AGB  &  J051113.88-673616.2 &  8.64 &  7.66 &  6.66 &  5.78 &  4.19 \\
  48   &C AGB?  &  J050721.64-674742.9 &  9.09 &  8.17 &  7.29 &  6.42 &  4.03 \\
  83   & C AGB  &  J050854.15-690046.2 &  8.18 &  7.22 &  6.33 &  5.43 &  3.83 \\
  87   & C AGB  &  J051019.61-694951.5 &  9.14 &  8.34 &  7.53 &  6.67 &  5.27 \\
  91   & C AGB  &  J050338.54-685312.8 &  8.94 &  7.67 &  6.65 &  5.69 &  3.87 \\
  95   & C AGB  &  J050959.97-695609.4 &  7.79 &  7.12 &  6.54 &  5.93 &  4.94 \\
 108   & C AGB  &  J051056.78-693530.4 &  7.82 &  7.15 &  6.50 &  5.71 &  4.33 \\
 138   & C AGB  &  J050137.87-712112.3 &  8.97 &  8.13 &  7.34 &  6.58 &  5.63 \\
 141   &   RSG  &  J050533.44-703346.8 &  7.27 &  7.26 &  6.97 &  6.42 &  3.98 \\
 196   & C AGB  &  J051200.77-703224.2 &  9.94 &  8.21 &  6.82 &  5.44 &  3.07 \\
 202   & C AGB  &  J051414.83-700409.7 & 10.04 &  8.96 &  8.02 &  7.04 &  5.37 \\
 216   &O AGB?  &  J051338.88-692108.1 &  7.72 &  7.04 &  6.41 &  5.63 &  4.12 \\
 218   & C AGB  &  J051316.38-684410.1 &  7.80 &  7.15 &  6.55 &  6.01 &  5.04 \\
 219   & C AGB  &  J051119.49-684227.7 &  8.82 &  7.73 &  6.86 &  6.01 &  4.46 \\
 220   & C AGB  &  J051232.01-691540.5 &  8.26 &  7.44 &  6.62 &  5.82 &  4.70 \\
 221   & C AGB  &  J051515.65-690033.7 &  9.18 &  8.35 &  7.55 &  6.68 &  5.14 \\
 223   & C AGB  &  J051251.00-693750.3 &  9.35 &  8.29 &  7.31 &  6.33 &  4.74 \\
 224   &   WR?  &  J051354.21-693146.6 &  7.55 &  7.47 &  7.14 &  6.38 &  4.60 \\
 225   & C AGB  &  J051437.87-681920.7 &  9.06 &  7.99 &  7.04 &  6.14 &  4.38 \\
 262   &  B[e]  &  J051352.94-672654.7 &  6.25 &  5.55 &  4.95 &  4.33 &  3.10 \\
 263   &   RSG  &  J051246.35-671937.9 &  7.16 &  6.88 &  6.56 &  5.95 &  3.56 \\
 264   &   RSG  &  J051449.73-672719.6 &  6.92 &  6.93 &  6.58 &  6.06 &  3.88 \\
 283   & O AGB  &  J051304.55-645140.2 &  8.29 &  7.67 &  7.13 &  6.17 &  3.65 \\
 307   & C AGB  &  J051856.19-674504.5 &  8.64 &  7.31 &  6.20 &  5.15 &  3.23 \\
 320   &  HII?  &  J051916.35-692027.9 & 11.94 &  9.92 &  8.17 &  6.57 &  3.97 \\
 321   &        &  J051903.69-692932.3 &  8.03 &  7.17 &  6.36 &  5.50 &  4.18 \\
 322   &        &  J051610.66-691441.1 &  9.82 &  8.43 &  7.27 &  6.16 &  4.35 \\
 323   &  B[e]  &  J051631.78-682209.5 &  8.23 &  7.48 &  6.80 &  5.94 &  3.93 \\
 325   & C AGB  &  J051637.68-692714.2 &  9.42 &  8.44 &  7.52 &  6.54 &  5.05 \\
 341   & C AGB  &  J052100.38-692055.3 &  9.81 &  8.64 &  7.71 &  6.83 &  5.06 \\
 344   &        &  J051722.54-692015.5 &  7.98 &  7.45 &  6.81 &  6.01 &  4.18 \\
 349   & C AGB  &  J051726.92-685458.6 & 10.36 &  9.10 &  8.03 &  7.10 &  5.22 \\
 357   & C AGB  &  J051622.61-695418.0 &  7.88 &  7.07 &  6.33 &  5.57 &  4.30 \\
 362   &   GMV  &  J051638.54-704541.3 &  6.87 &  7.14 &  6.84 &  6.29 &  4.20 \\
 420   & C AGB  &  J052425.04-714901.9 &  8.88 &  8.04 &  7.29 &  6.53 &  5.21 \\
 435   &C AGB?  &  J052430.79-695353.3 &  8.07 &  7.06 &  6.06 &  5.12 &  3.48 \\
 436   &C AGB?  &  J052240.87-701025.1 &  9.17 &  8.16 &  7.15 &  6.20 &  4.66 \\
 438   & C AGB  &  J052519.50-710402.4 &  8.65 &  7.86 &  7.13 &  6.40 &  5.46 \\
 441   & C AGB  &  J052438.62-702356.9 &  9.84 &  8.69 &  7.61 &  6.59 &  4.74 \\
 461   &  HII?  &  J052419.26-693849.2 &  6.34 &  6.34 &  6.02 &  5.42 &  3.33 \\
 464   &  HII?  &  J052413.34-682958.7 &  9.13 &  8.35 &  7.48 &  6.40 &  2.32 \\
 466   &        &  J052150.59-682958.2 & 10.11 &  8.61 &  7.36 &  6.27 &  4.29 \\
 467   &  HII?  &  J052413.41-694112.8 & 10.81 &  8.91 &  7.29 &  5.81 &  3.31 \\
 506   &  RSG?  &  J052315.45-675941.0 &  8.30 &  8.36 &  8.09 &  7.83 &  5.38 \\
 507   &  HII?  &  J052458.46-675838.3 & 10.08 &  8.63 &  7.40 &  6.31 &  4.20 \\
 529   &   RSG  &  J052343.60-654159.8 &  7.32 &  7.38 &  7.08 &  6.46 &  3.86 \\
 549   &   RSG  &  J052611.32-661211.1 &  7.11 &  7.00 &  6.60 &  5.99 &  4.43 \\
 551   &   RSG  &  J052616.06-660658.9 &  7.70 &  7.57 &  7.19 &  6.57 &  4.20 \\
 558   &   RSG  &  J053020.93-672005.4 &  7.01 &  6.78 &  6.43 &  5.77 &  3.82 \\
 560   & C AGB  &  J053003.88-664924.2 &  9.31 &  8.21 &  7.25 &  6.32 &  4.66 \\
 561   &    PN  &  J052917.44-671329.9 &  7.64 &  6.73 &  5.89 &  5.09 &  3.52 \\
 562   &  HII?  &  J052717.79-662205.7 &  8.47 &  7.77 &  7.10 &  6.16 &  3.46 \\
 585   & B[e]?  &  J052653.11-685000.3 &  6.21 &  6.12 &  5.73 &  4.80 &  2.58 \\
 587   &   RSG  &  J053104.14-691902.9 &  6.84 &  6.86 &  6.46 &  5.74 &  4.09 \\
 588   &   RSG  &  J052747.49-691320.5 &  7.30 &  7.27 &  6.84 &  6.08 &  4.06 \\
 589   &   RSG  &  J052634.76-685140.0 &  6.86 &  6.92 &  6.55 &  6.07 &  4.14 \\
 590   &   RSG  &  J052927.59-690850.4 &  6.86 &  6.82 &  6.51 &  5.88 &  4.01 \\
 591   &  RSG?  &  J053100.38-691047.5 &  7.53 &  7.31 &  6.98 &  6.24 &  4.37 \\
 592   &   RSG  &  J052627.38-691055.9 &  7.36 &  7.44 &  7.14 &  6.53 &  4.13 \\
 593   &   RSG  &  J052828.86-680707.9 &  6.86 &  6.82 &  6.58 &  6.14 &  3.97 \\
 594   &C AGB?  &  J052648.83-692335.3 &  8.84 &  7.79 &  6.84 &  5.94 &  4.23 \\
 597   &   RSG  &  J052942.19-685717.4 &  6.65 &  6.81 &  6.53 &  6.15 &  4.48 \\
 599   &  HII?  &  J052627.16-684842.1 &  \nodata & \nodata & 10.15 & \nodata & 1.46 \\
 601   & C AGB  &  J052650.81-693136.9 &  8.89 &  8.15 &  7.37 &  6.55 &  5.39 \\
 635   & C AGB  &  J052724.08-693944.9 &  8.24 &  7.01 &  5.90 &  4.88 &  3.06 \\
 639   &  HII?  &  J053051.75-694327.9 &  8.71 &  7.50 &  6.51 &  5.56 &  3.76 \\
 642   & O AGB  &  J052848.15-710229.0 &  7.98 &  7.69 &  7.24 &  6.05 &  3.46 \\
 643   &C AGB?  &  J052804.03-695308.2 &  8.31 &  7.43 &  6.63 &  5.86 &  4.40 \\
 644   & C AGB  &  J052922.85-700646.3 &  8.82 &  7.79 &  6.88 &  6.01 &  4.12 \\
 645   & C AGB  &  J053104.75-710119.0 &  8.02 &  7.45 &  6.78 &  6.19 &  5.30 \\
 653   &C AGB?  &  J052924.60-695513.8 &  7.44 &  6.67 &  5.99 &  5.03 &  2.09 \\
 661   & C AGB  &  J052703.96-710859.0 &  8.85 &  8.17 &  7.51 &  6.83 &  5.53 \\
 689   & C AGB  &  J052737.82-712436.6 &  8.56 &  7.72 &  6.99 &  6.24 &  5.24 \\
 692   & C AGB  &  J052846.58-711912.5 &  9.34 &  8.28 &  7.35 &  6.49 &  4.84 \\
 716   &   GMV  &  J053301.04-715741.6 & \nodata & \nodata &  4.58 &  4.41 &  3.62 \\
 720   &C AGB?  &  J053634.19-714415.8 &  9.14 &  8.23 &  7.35 &  6.56 &  5.16 \\
 733   & C AGB  &  J053415.93-702252.6 &  8.78 &  7.50 &  6.48 &  5.49 &  3.88 \\
 745   &C AGB?  &  J053301.43-701322.5 &  8.97 &  8.00 &  7.07 &  6.29 &  5.18 \\
 768   & C AGB  &  J053140.85-693919.5 &  8.30 &  7.42 &  6.56 &  5.76 &  4.38 \\
 771   & C AGB  &  J053238.58-682522.3 &  8.89 &  7.79 &  6.87 &  5.92 &  2.23 \\
 772   &  HII?  &  J053319.74-694144.8 &  9.85 &  8.31 &  7.14 &  6.05 &  4.14 \\
 773   &   WR?  &  J053541.10-691159.6 &  7.35 &  7.21 &  6.81 &  6.21 &  4.89 \\
 774   & C AGB  &  J052623.10-691120.3 &  9.71 &  8.63 &  7.61 &  6.61 &  4.84 \\
 775   & C AGB  &  J053256.15-681248.8 &  7.68 &  7.07 &  6.62 &  6.19 &  6.15 \\
 805   &   RSG  &  J053235.61-675509.3 &  6.71 &  6.40 &  6.04 &  5.46 &  3.04 \\
 807   & O AGB  &  J053237.17-670656.2 &  7.99 &  7.29 &  6.62 &  5.66 &  3.00 \\
 809   &   RSG  &  J053326.80-670413.4 &  7.27 &  7.21 &  6.79 &  6.09 &  4.02 \\
 810   &   RSG  &  J053020.63-665301.8 &  7.78 &  7.97 &  7.63 &  7.08 &  5.38 \\
 811   & C AGB  &  J053251.36-670651.7 &  7.22 &  6.43 &  5.74 &  4.75 &  1.89 \\
 813   &   RSG  &  J053307.62-664805.5 &  7.88 &  7.57 &  7.19 &  6.31 &  4.40 \\
 815   &   RSG  &  J053514.07-674355.7 &  7.40 &  7.15 &  6.87 &  6.31 &  3.91 \\
 836   &   HII  &  J053231.94-662715.2 & \nodata & 11.93 &  9.58 &  7.81 &  2.64 \\
 837   &        &  J053147.40-660340.7 &  7.74 &  7.29 &  6.82 &  5.95 &  3.15 \\
 839   &   RSG  &  J053136.82-663007.8 &  6.95 &  6.93 &  6.63 &  6.21 &  4.16 \\
 840   & C AGB  &  J052953.60-651456.6 &  8.69 &  7.88 &  7.07 &  6.29 &  5.09 \\
 869   &   RSG  &  J053625.46-665538.6 &  6.89 &  6.94 &  6.70 &  6.11 &  3.56 \\
 870   &   RSG  &  J053528.28-665602.3 &  6.89 &  6.97 &  6.70 &  6.16 &  4.01 \\
 887   &C/O AGB &  J054013.32-692246.4 &  6.48 &  5.84 &  5.28 &  4.51 &  2.89 \\
 890   &  B[e]  &  J053625.84-692255.7 &  6.77 &  6.07 &  5.47 &  4.70 &  2.01 \\
 891   &   RSG  &  J053555.22-690959.4 &  6.82 &  6.77 &  6.33 &  5.57 &  3.68 \\
 892   &   RSG  &  J053745.05-692048.9 &  7.12 &  6.99 &  6.63 &  5.99 &  3.69 \\
 893   &   RSG  &  J054059.15-691836.1 &  7.19 &  6.90 &  6.63 &  6.07 &  3.68 \\
 897   &   RSG  &  J054043.72-692158.1 &  7.03 &  7.02 &  6.64 &  6.07 &  4.62 \\
 906   &  HII?  &  J053630.80-691817.3 & 11.25 &  9.94 &  8.56 &  6.58 &  1.87 \\
 937   & C AGB  &  J054036.06-695249.9 &  8.60 &  7.60 &  6.72 &  5.83 &  4.29 \\
 939   &   RSG  &  J054048.48-693336.0 &  7.16 &  7.01 &  6.58 &  5.76 &  3.45 \\
 940   &  HII?  &  J053931.15-701216.8 & 10.53 &  9.10 &  7.82 &  6.61 &  1.69 \\
 941   & C AGB  &  J053901.71-700843.0 &  8.70 &  7.85 &  7.06 &  6.33 &  5.19 \\
 943   &   RSG  &  J054110.59-693803.9 &  7.12 &  7.18 &  6.91 &  6.50 &  4.42 \\
 981   &C AGB?  &  J054243.27-711133.8 &  9.28 &  8.15 &  7.17 &  6.29 &  4.63 \\
1048   &   GMV  &  J043646.82-701841.4 & \nodata &  5.53 &  5.34 &  5.12 &  4.63 \\
1072   & O AGB  &  J044028.48-695513.8 &  7.10 &  6.70 &  6.27 &  5.17 &  2.56 \\
1117   & O AGB  &  J044941.42-683751.4 &  6.56 &  6.27 &  5.84 &  4.72 &  2.53 \\
1119   & C AGB  &  J044738.83-692117.0 &  8.81 &  7.99 &  7.21 &  6.53 &  5.48 \\
1120   & C AGB  &  J044716.06-682425.9 &  7.82 &  7.11 &  6.42 &  5.76 &  4.67 \\
1130   & C AGB  &  J044918.46-695314.4 &  7.49 &  6.89 &  6.25 &  5.59 &  3.95 \\
1132   &   RSG  &  J044922.45-692434.3 &  7.29 &  7.24 &  6.93 &  6.36 &  3.91 \\
1171   &O AGB?  &  J045410.03-695558.2 &  8.40 &  7.81 &  7.21 &  6.36 &  3.18 \\
1173   &C AGB?  &  J045150.90-703832.0 &  7.88 &  6.85 &  6.00 &  5.14 &  3.59 \\
1184   &   HII  &  J045245.65-691149.5 &  7.42 &  6.48 &  5.61 &  4.50 &  1.05 \\
1186   &  HII?  &  J045401.18-691152.6 & \nodata & \nodata &  9.27 & \nodata &  1.27 \\
1189   &   RSG  &  J045503.06-692912.9 &  6.82 &  6.58 &  6.24 &  5.52 &  2.89 \\
1190   &O AGB?  &  J045129.03-685749.8 &  7.37 &  7.02 &  6.54 &  5.48 &  3.15 \\
1191   &   RSG  &  J045330.90-691749.5 &  7.48 &  7.45 &  7.12 &  6.46 &  1.86 \\
1192   & O AGB  &  J045040.43-691731.8 &  7.09 &  6.82 &  6.39 &  5.22 &  2.68 \\
1193   & C AGB  &  J045023.41-693756.6 &  8.70 &  7.79 &  7.03 &  6.30 &  4.84 \\
1204   &   RSG  &  J045516.02-691912.0 &  7.11 &  7.19 &  6.91 &  6.37 &  5.01 \\
1207   &  HII?  &  J045506.52-691708.5 &  9.41 &  8.61 &  7.82 &  6.72 &  2.06 \\
1225   &        &  J045747.94-662845.0 &  8.55 &  7.66 &  6.91 &  5.76 &  2.16 \\
1227   &        &  J045405.75-664506.9 &  9.40 &  8.48 &  7.70 &  6.53 &  2.62 \\
1229   &  HII?  &  J045640.75-663230.4 & 10.46 &  9.23 &  8.07 &  6.83 &  2.36 \\
1278   & C AGB  &  J050104.36-661240.4 &  8.58 &  7.51 &  6.66 &  5.85 &  4.50 \\
1280   & O AGB  &  J050018.99-670758.0 &  8.52 &  8.06 &  7.53 &  6.59 &  3.92 \\
1282   & C AGB  &  J050100.83-673523.6 &  8.83 &  7.99 &  7.14 &  6.28 &  5.00 \\
1298   & C AGB  &  J045632.14-685251.0 &  7.89 &  7.02 &  6.27 &  5.52 &  4.43 \\
1302   &  HII?  &  J045845.98-682037.7 & 10.74 &  8.88 &  7.40 &  6.08 &  3.66 \\
1326   &  B[e]  &  J045647.07-695024.8 &  7.18 &  6.44 &  5.80 &  4.82 &  2.36 \\
1328   &   RSG  &  J045743.26-700850.3 &  6.70 &  6.72 &  6.44 &  5.79 &  3.28 \\
1329   &   RSG  &  J045534.83-692655.6 &  6.85 &  6.94 &  6.55 &  5.94 &  3.89 \\
1330   &   RSG  &  J045521.58-694716.7 &  7.40 &  7.22 &  6.92 &  6.33 &  3.80 \\
1360   &C AGB?  &  J045614.98-694047.9 &  9.43 &  8.31 &  7.33 &  6.42 &  4.60 \\
1378   & C AGB  &  J054650.87-712803.4 &  9.58 &  8.42 &  7.47 &  6.59 &  4.82 \\
1379   &        &  J054705.56-703433.4 &  9.46 &  8.13 &  7.05 &  6.01 &  4.05 \\
1383   &C AGB?  &  J054413.64-694418.2 &  9.49 &  8.07 &  6.96 &  5.91 &  3.92 \\
1384   & C AGB  &  J054336.02-701035.0 &  9.06 &  8.22 &  7.45 &  6.79 &  5.34 \\
1400   & C AGB  &  J054020.57-661444.3 &  9.18 &  8.11 &  7.16 &  6.28 &  4.61 \\
1429   &   RSG  &  J054413.78-661644.5 &  7.04 &  7.02 &  6.74 &  6.41 &  4.65 \\
1436   &C AGB?  &  J054330.31-692446.6 &  9.85 &  8.78 &  7.81 &  6.41 &  2.57 \\
1453   &C AGB?  &  J054956.51-705311.8 &  9.94 &  8.37 &  7.08 &  5.81 &  3.61 \\
1456   & C AGB  &  J055303.36-703316.9 &  7.72 &  6.95 &  6.25 &  5.58 &  4.56 \\
1471   & C AGB  &  J054443.79-694831.9 &  9.85 &  8.71 &  7.76 &  6.73 &  5.16 \\
1492   & C AGB  &  J054908.87-713206.9 &  7.89 &  7.43 &  6.91 &  6.39 &  6.06 \\
1524   & O AGB  &  J055521.04-700002.8 &  8.00 &  7.44 &  6.88 &  5.67 &  2.90 \\
1546   & C AGB  &  J055026.05-674946.1 &  8.35 &  7.45 &  6.60 &  5.76 &  4.40 \\
1589   & O AGB  &  J055651.49-682726.6 &  7.22 &  6.96 &  6.64 &  5.95 &  3.80 \\
1651   & C AGB  &  J060245.12-672243.2 &  8.54 &  7.30 &  6.27 &  5.26 &  3.43 \\
1652   & C AGB  &  J060231.07-671246.9 &  7.97 &  6.81 &  5.82 &  4.89 &  3.22 \\
1653   & C AGB  &  J055959.36-674156.8 &  9.83 &  8.78 &  7.88 &  6.91 &  5.38 \\
1780   & C AGB  &  J053301.74-682358.4 & 10.22 &  8.55 &  7.36 &  6.11 &  3.77 \\
1794   &  HII?  &  J054044.00-692554.5 & 11.69 & 10.84 &  9.53 & \nodata &  3.49 \\
1797   &C/O AGB &  J054217.17-703220.3 &  8.85 &  8.25 &  7.56 &  6.81 &  5.56 \\
\enddata

\tablenotetext{a}{Object classification, from col.\ 11 of Table 1.}

\end{deluxetable}

\begin{deluxetable}{lcc}
\rotate
\tabletypesize{\scriptsize}
\tablewidth{0pt}
\tablecolumns{3}
\tablecaption{Luminous LMC Mid-IR Sources: IRAC/MIPS Color Classification Criteria}
\tablehead{
\colhead{Class} &
\multicolumn{2}{c}{Criteria}  \\
}

\startdata
RSG              &  [-2.30 $\times$ ([5.8]$-$[8.0]) + 2.42] $\le$ [8.0]$-$[24] $\le$ [-2.30 $\times$ ([5.8]$-$[8.0]) + 4.22] & [0.49 $\times$ ([8.0]$-$[24]) $-0.50$] $\le$ [5.8]$-$[8.0] $\le$[0.49 $\times$ ([8.0]$-$[24]) $-0.07$] \\
O AGB            &  2.5 $\le$ [8.0]$-$[24] $\le$ 2.85 & 0.925 $\le$ [5.8]$-$[8.0] $\le$ 1.25 \\
C AGB            &  [-2.30 $\times$ ([5.8]$-$[8.0]) + 2.09] $\le$ [8.0]$-$[24] $\le$ [-2.30 $\times$ ([5.8]$-$[8.0]) + 4.22] & [0.49 $\times$ ([8.0]$-$[24]) $-0.06$] $\le$ [5.8]$-$[8.0] $\le$ [0.49 $\times$ ([8.0]$-$[24]) $+0.34$]
 \\
\enddata

\end{deluxetable}

\end{document}